\pdfoutput=1
\documentclass[aip]{revtex4-1}
\usepackage{graphicx}
\usepackage{dcolumn}
\usepackage{bm}

\usepackage[utf8]{inputenc}
\usepackage[T1]{fontenc}
\usepackage{mathptmx}
\usepackage{amsmath}
\usepackage{etoolbox}
\usepackage{multirow}
\usepackage[hidelinks]{hyperref}
\usepackage[capitalise]{cleveref}
\usepackage{caption}
\usepackage{subcaption}
\usepackage{textgreek}
\usepackage{nicefrac}
\usepackage{booktabs}
\usepackage{lipsum}
\begin{document}


\title{Characterising the Surface Resistance of Laser-Treated LHC Beam Screens with the Shielded Pair Method}



\author{Patrick Krkoti\'c}
\email[email address: ]{patrick.krkotic@cern.ch}
\author{Elena Bez}
\altaffiliation{University of Leipzig, Linnestraße 5, 04103 Leipzig, Germany}
\author{Krist\'of Brunner}
\author{Sergio Calatroni}
\email[email address: ]{sergio.calatroni@cern.ch}
\author{Maleesh Dissanayake}
\altaffiliation{Paris-Saclay University, 3 rue Joliot Curie, 91190 Gif-sur-Yvette}
\author{Marcel Himmerlich}
\email[email address: ]{marcel.himmerlich@cern.ch}
\author{Ana Karen Reascos Portilla}
\author{Mauro Taborelli}
\author{Matthew James Watkins}
\affiliation{CERN, European Organization for Nuclear Research, 1211 Geneva 23, Switzerland}

\date{\today}

\begin{abstract}
The presence of strong electron clouds in the quadrupole magnetic field regions of the Large Hadron Collider (LHC) leads to considerable heating that poses challenges for the cryogenic cooling system, and under certain conditions to proton beam quality deterioration. Research is being conducted on laser-treated inner beam screen surfaces for the upgraded High-Luminosity LHC to mitigate this issue. Laser-induced surface structuring, a technique that effectively roughens surfaces, has been shown to reduce secondary electron emission; an essential factor in controlling electron cloud formation. Conversely, the resulting surface roughening also alters the material’s surface impedance, potentially impacting beam stability and increasing beam-induced resistive wall heating. Different laser treatment patterns have been applied to LHC beam screens to estimate this potential impact and assessed for their microwave responses. 
\end{abstract}


\maketitle

\section{\label{sec:1} Introduction}
With the High-Luminosity (HL) upgrade of the Large Hadron Collider (LHC) at CERN, a significant increase in electron cloud formation is anticipated within the vacuum system. This phenomenon affects both the heat load on the cryogenic cooling system and the quality of the circulating proton beam, making mitigation strategies essential.

Electron clouds typically form in regions with strong magnetic fields, such as those in dipole and quadrupole magnets, where magnetic fields trap electrons along field lines, increasing their density and impact \cite{Skripka}. Additionally, the inner surface properties of the beam screen (BS) play a critical role in determining local electron density. The LHC BS consists of a 75\,\textmu m thick layer of oxygen-free electronic (OFE) copper co-laminated onto stainless steel, designed to shield the magnets operating at 1.9\,K from excessive heat load and synchrotron radiation.

The secondary electron yield (SEY) of copper, which is one of the important parameters that determine the probability of electron cloud formation, depends heavily on its oxidation state and surface impurities \cite{HimmerlichMarcel}. Prior research has shown that focused ultrashort-pulsed laser treatments can significantly reduce the SEY by creating micro- and nano-structures tailored to processing parameters such as laser power, line spacing, and scanning speed \cite{Bez2023a}.

For the HL-LHC, advanced laser treatment technology has been developed to selectively treat the BS surface, particularly within the quadrupole magnet regions, such as the Q5 standalone magnets near the CMS and ATLAS detectors during the third long shutdown \cite{BezSelective}. Selective treatment allows for precise targeting of areas prone to electron cloud formation \cite{Sitko}. However, surface modifications, such as roughening, can influence the effective surface resistance of the BS, potentially impacting beam stability and increasing resistive wall heating. This necessitates experimental validation of laser treatments to ensure that they mitigate electron clouds without introducing detrimental side effects. While a number of studies have determined the surface resistance of flat laser-treated samples with processed surface areas ranging from 1\,cm$^2$ to 80\,cm$^2$, the surface resistance of a laser-treated beam screen has never been directly measured.

To address these challenges, Brunner \textit{et al.} \cite{KristofIPAC, KristofPhD, KristofMDPI} developed a radio frequency (RF) characterisation setup for LHC BSs, adaptable to new designs for future colliders. Covering a frequency range of 400 to 1600\,MHz and temperatures from 4.2\,K to room temperature, the setup uses the non-destructive shielded pair method \cite{Smith}, where the BS forms a resonant structure. It has been successfully applied to standard LHC BSs and amorphous carbon-coated variants, demonstrating minimal surface resistance impact for coatings like a 200\,nm titanium underlayer with a 50\,nm amorphous carbon layer \cite{KristofMDPI}. This setup is now employed to evaluate the impact of laser treatments on the surface resistance of LHC BSs.

\begin{figure}[b!]
    \centering
    \includegraphics[width=\linewidth]{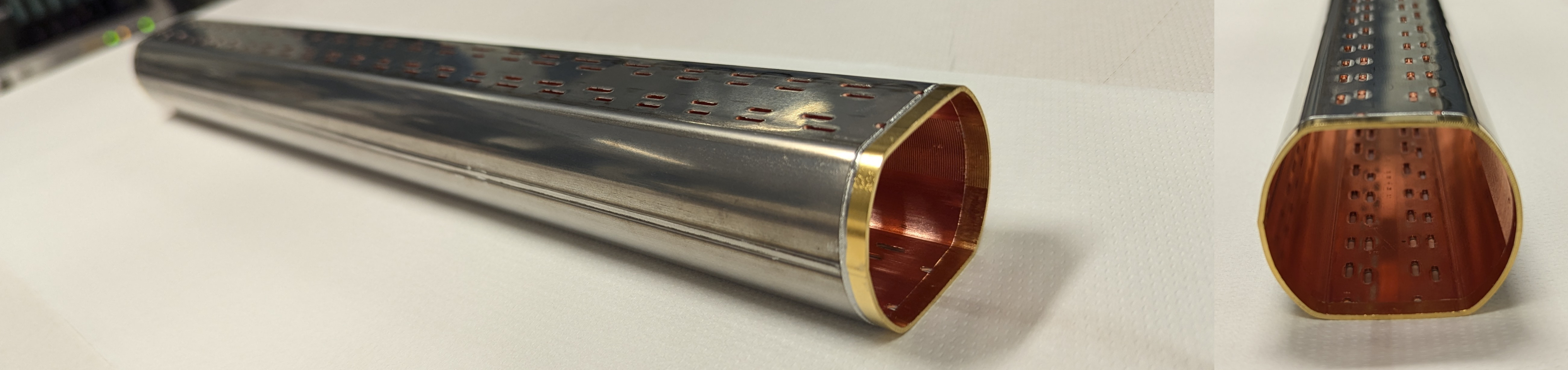}
    \caption{Photograph of a standard LHC beam screen sample with gold contact rings added on each extremity.}
    \label{fig:LHCBS}
\end{figure}

\section{LHC Beam Screen Laser Treatment \label{sec:2}}
Four standard LHC beam screens (type 50A), each 400\,mm in length and equipped with gold contact rings at both ends were produced as shown in \Cref{fig:LHCBS}. Before further treatment, the BSs were wet-chemically degreased using a commercial detergent (DP 17.40 SUP from NGL) and then rinsed with deionised water \cite{BezSelective}. Three of the BSs underwent various laser treatments, which will be detailed later, while one remained untreated to serve as a reference.

\begin{figure}[t!]
     \centering
     \begin{subfigure}[b]{0.32\textwidth}
         \centering
         \includegraphics[width=\textwidth]{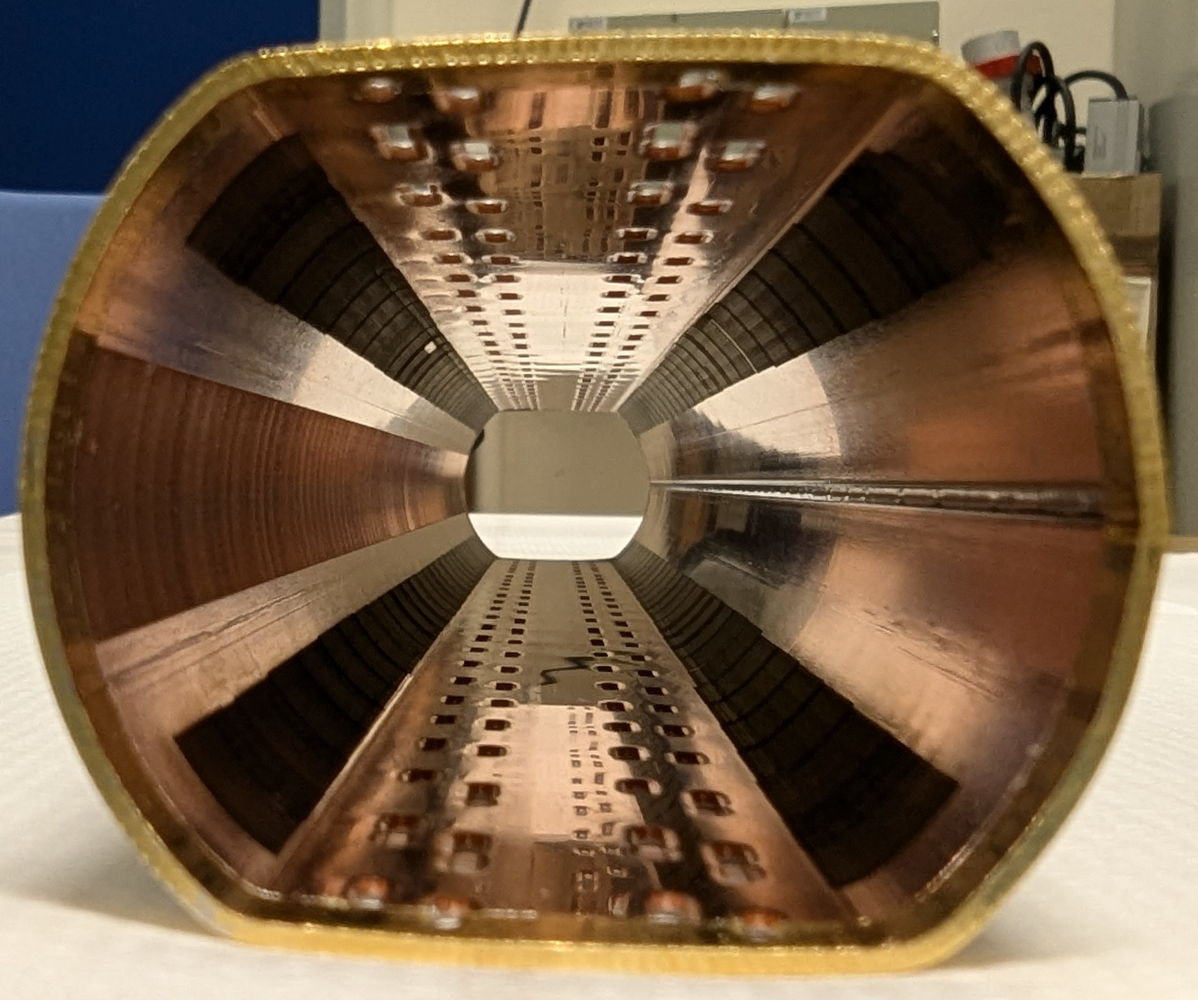}
         \includegraphics[width=\textwidth]{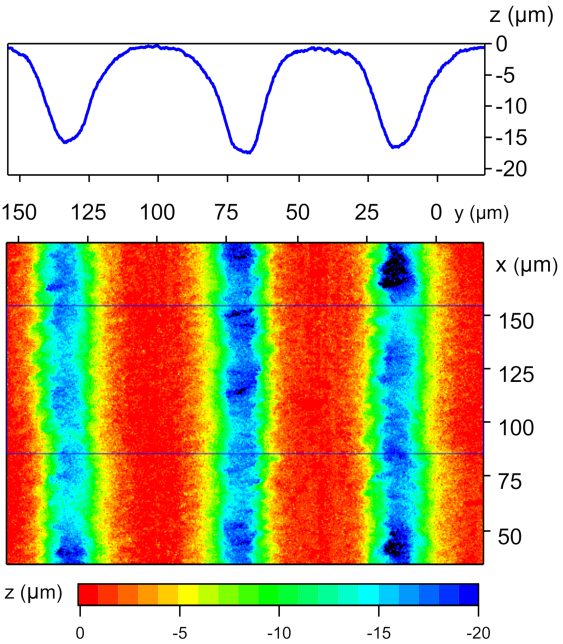}
         \caption{4 $\times$ 20$^\circ$ longline}
         \label{fig:Samples1}
     \end{subfigure}
     \hfill
     \begin{subfigure}[b]{0.32\textwidth}
         \centering
         \includegraphics[width=\textwidth]{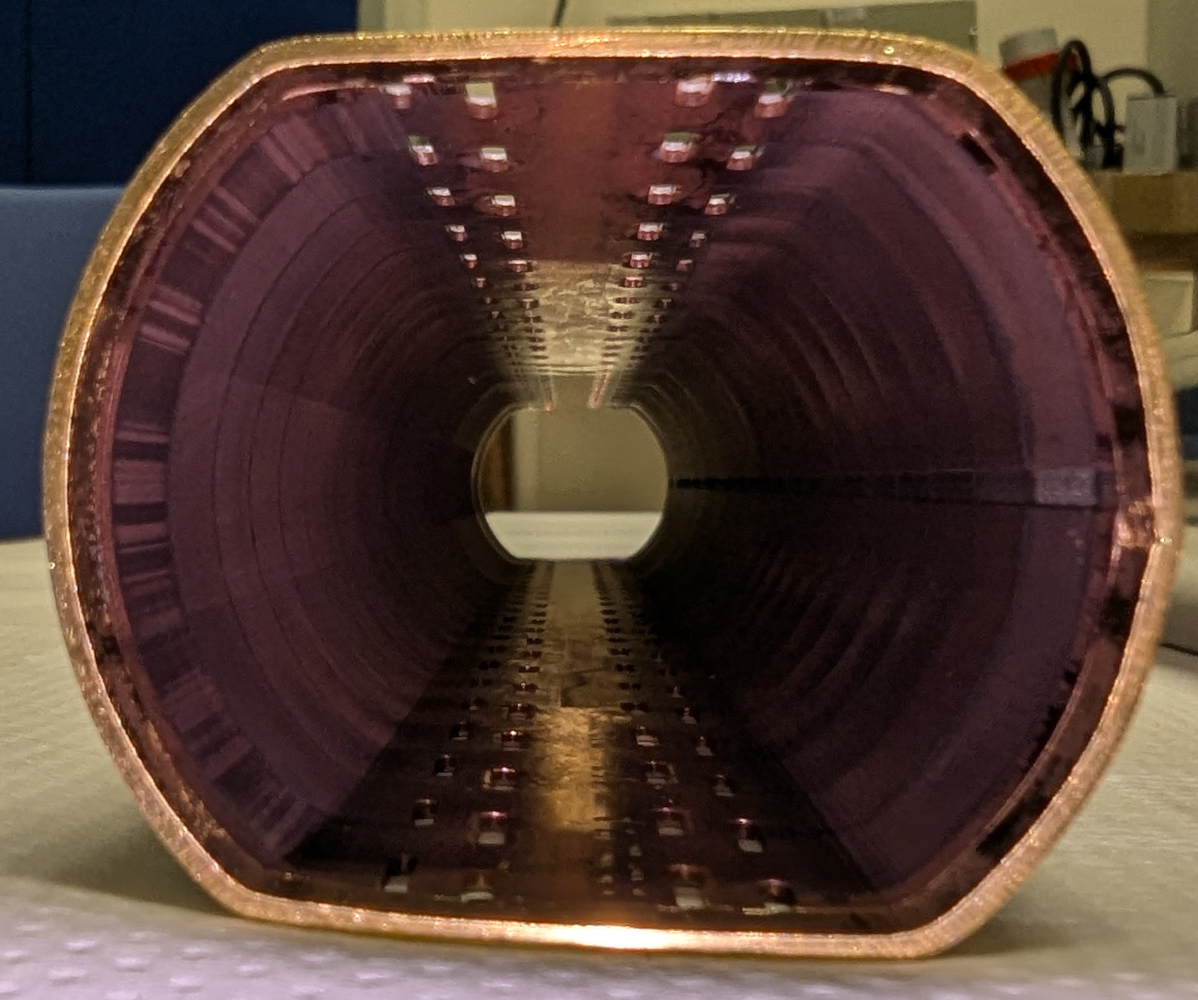}
        \includegraphics[width=\textwidth]{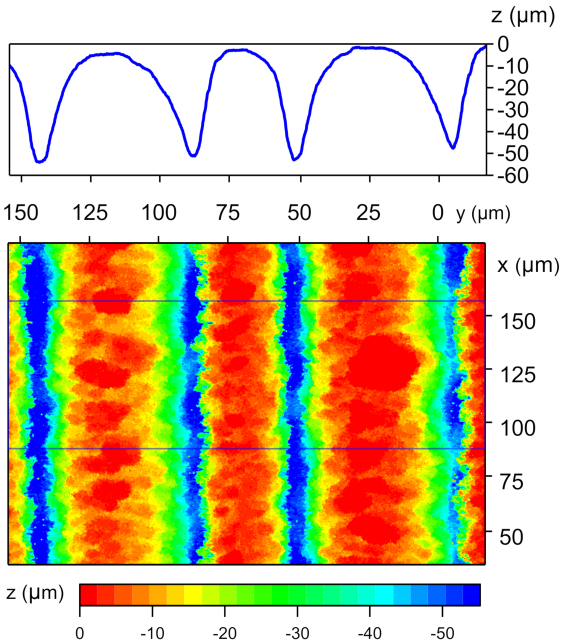}
         \caption{360$^\circ$ longline}
         \label{fig:Samples2}
     \end{subfigure}
     \hfill
     \begin{subfigure}[b]{0.32\textwidth}
         \centering
         \includegraphics[width=\textwidth]{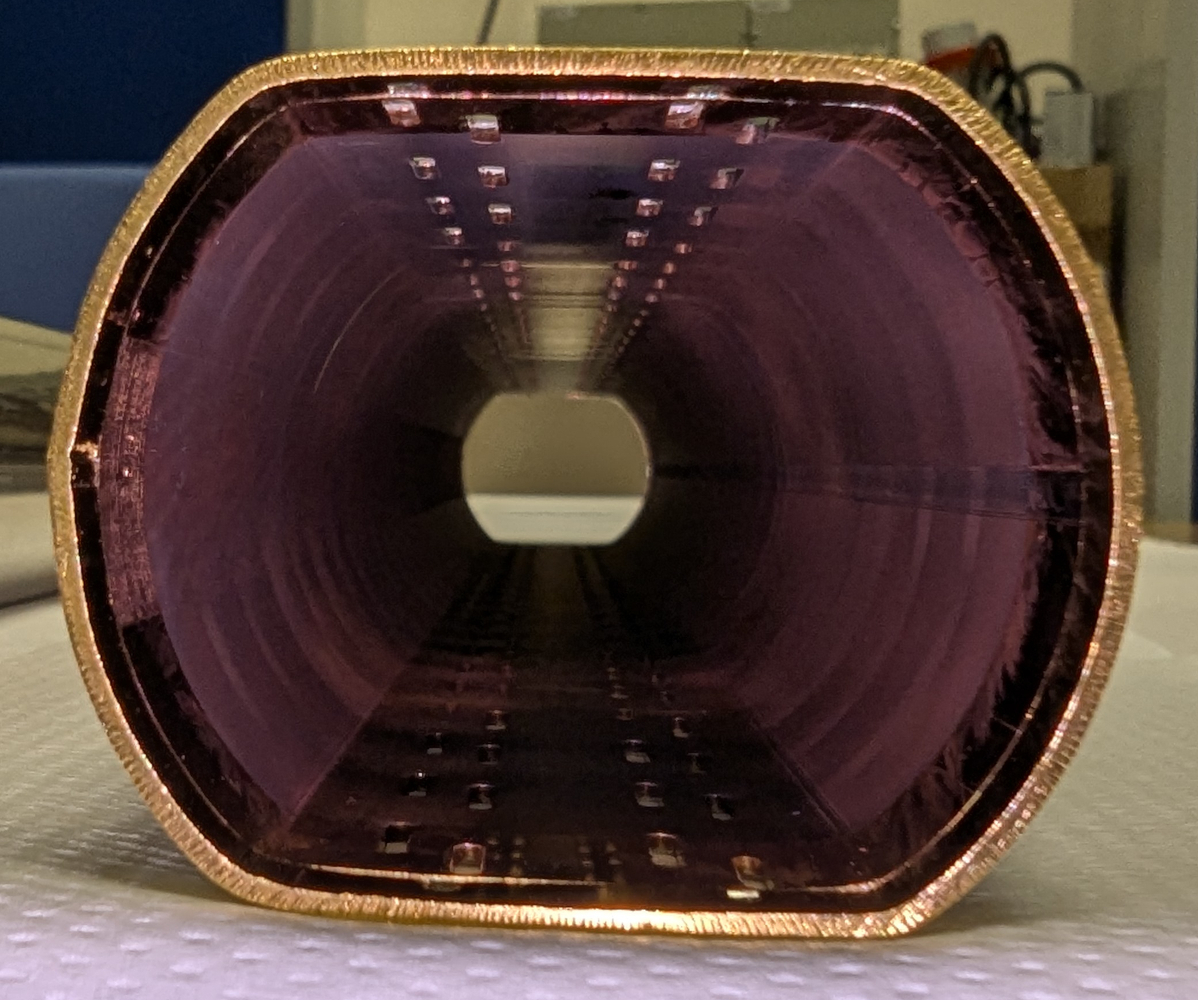}
         \includegraphics[width=\textwidth]{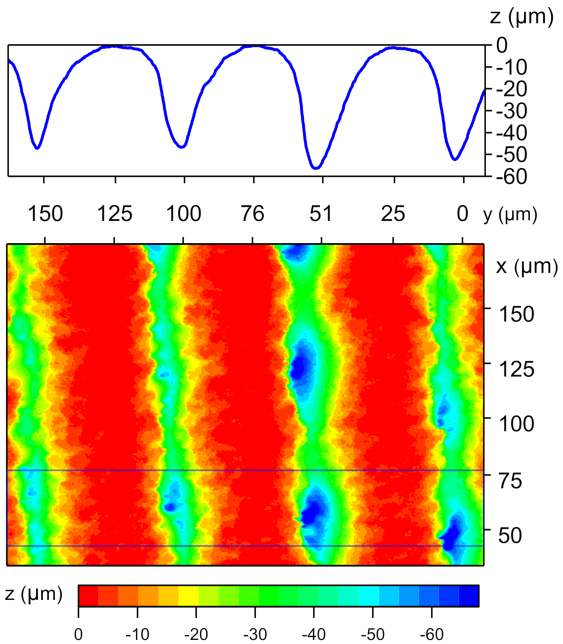}
         \caption{360$^\circ$ spiral}
         \label{fig:Samples3}
     \end{subfigure}
     \caption{Laser-patterned LHC Beam screen samples and resulting trench profiles. Trench profiles taken from \cite{BezPhD}.}
        \label{fig:Samples}
\end{figure}

The BSs underwent laser treatment using three patterns: 4 $\times$ 20$^\circ$ longline, 360$^\circ$ longline, and 360$^\circ$ spiral, as shown in \Cref{fig:Samples}. The treatment was performed with a linearly polarised, pulsed infrared laser beam (1\,ps, 1030\,nm) operating at a repetition rate of 500\,kHz, as described in Bez \textit{et al.} \cite{BezSelective} and Bez \cite{BezPhD}. The laser beam featured a Gaussian intensity profile, with the focused spot on the curved BS part measured at approximately 55\,\textmu m in diameter, and the spacing between scan lines set to about 50\,\textmu m.

The treatments were conducted at atmospheric pressure, with an additional nitrogen (N$_2$) flow of 5~\nicefrac{L}{min} to prevent oxidation during the process. For all three treated samples, particulates generated during the laser treatment were removed by exposing the surfaces to a stream of compressed gaseous N$_2$, effectively dislodging loose particles.

For the 4 $\times$ 20$^\circ$ longline treatment, only the four corners of the BS were processed, resulting in a total coverage of approximately 22\%. This treatment used an average laser power of 6.5\,W and a scanning speed of 20\,\nicefrac{mm}{s} to create grooves parallel to the BS's longitudinal axis. Due to the challenges of accessing the treated areas inside the BS, equivalent test samples were prepared for further analysis. One such analysis involved measuring the groove structure, as shown in \Cref{fig:Samples1}, which revealed groove depths varying between 17-22\,\textmu m.

The 360$^\circ$ longline treatment (\Cref{fig:Samples2}) extended the 4 $\times$ 20$^\circ$ longline treatment to cover the entire BS surface. This process utilised a higher average laser power of 8.5\,W and a slower scanning speed of 15\,\nicefrac{mm}{s}. Profiling results indicated that the trench depth in this sample was roughly double that of the 4 $\times$ 20$^\circ$ longline, ranging from 38-57\,\textmu m.

Lastly, the 360$^\circ$ spiral treatment (\Cref{fig:Samples3}) also processed the entire BS surface but used scanning lines oriented perpendicular to the longitudinal axis. The laser parameters for this treatment were the same as those for the 360$^\circ$ longline treatment, with an average power of 8.5\,W and a scanning speed of 15\,\nicefrac{mm}{s}. However, the profiling results showed trench depths ranging from 57-67\,\textmu m, which were deeper and exhibited less variation compared to the 360$^\circ$ longline treatment.

\begin{figure}[h!]
    \centering
     \includegraphics[width=0.9\textwidth]{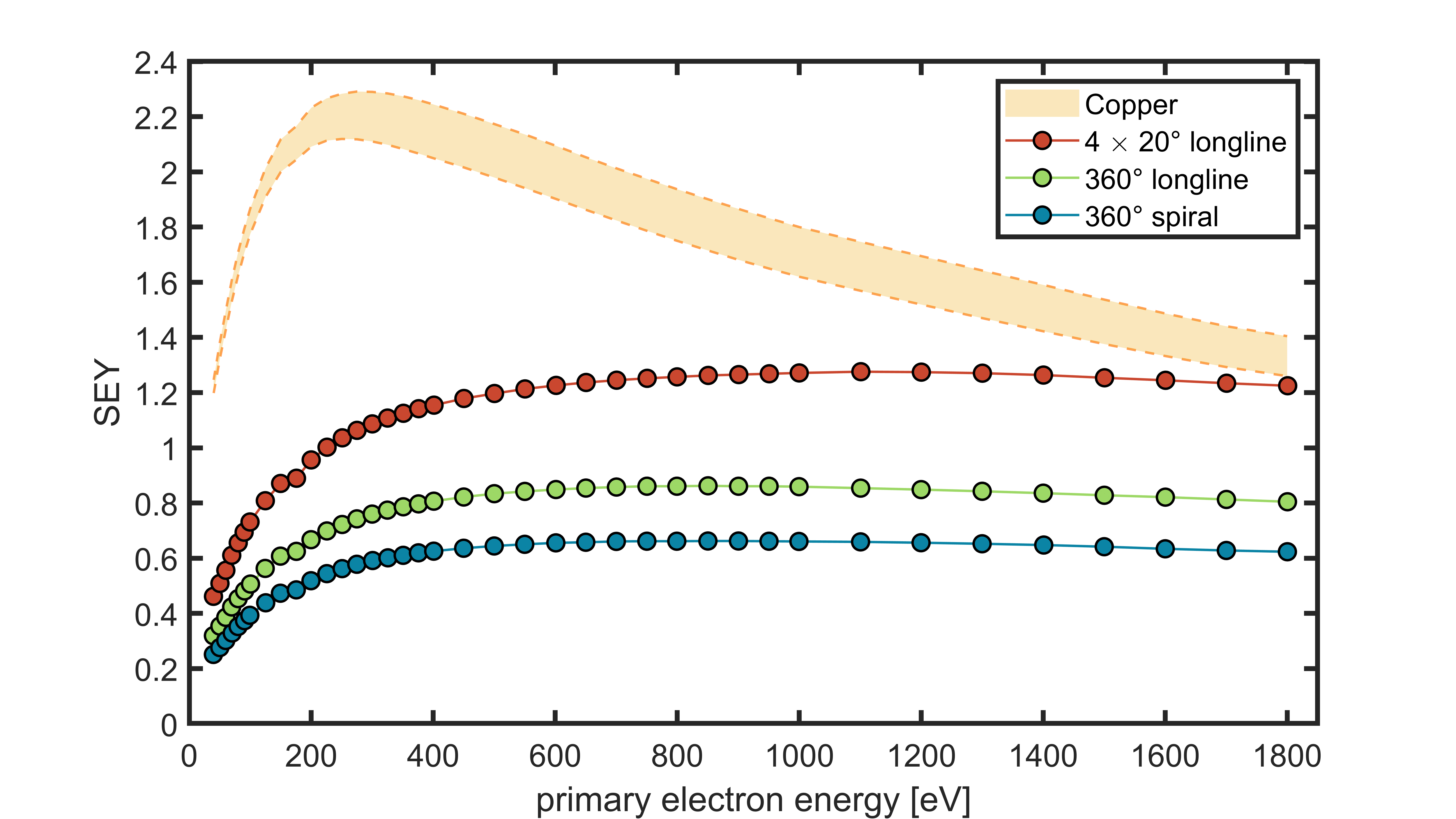}
    \caption{Secondary electron yield in the range between 50 and 1800\,eV for the three laser-patterned samples. Data taken from Bez \cite{BezPhD}.}
    \label{fig:SEY}
\end{figure}

The differences in the laser parameters are also reflected in the SEY for primary electron energies between 50 and 1800\,eV. SEY measurements were conducted on the test samples using the alternating sample bias method \cite{PhysRevAccelBeams.22.083101, Bez2023a}. \Cref{fig:SEY} presents the measurement results, showing that the SEY curves correlate with the groove depths. While the 360$^\circ$ spiral and 360$^\circ$ longline samples exhibit SEY values below 1, the maximum SEY for the 4 $\times$ 20$^\circ$ longline sample is nearly 1.2. For comparison, the typical SEY variation range for degreased copper is included.

\section{Methods and Materials \label{sec:3}}

\begin{figure}[t!]
     \centering
     \begin{subfigure}[b]{0.32\textwidth}
         \centering
         \includegraphics[width=\textwidth]{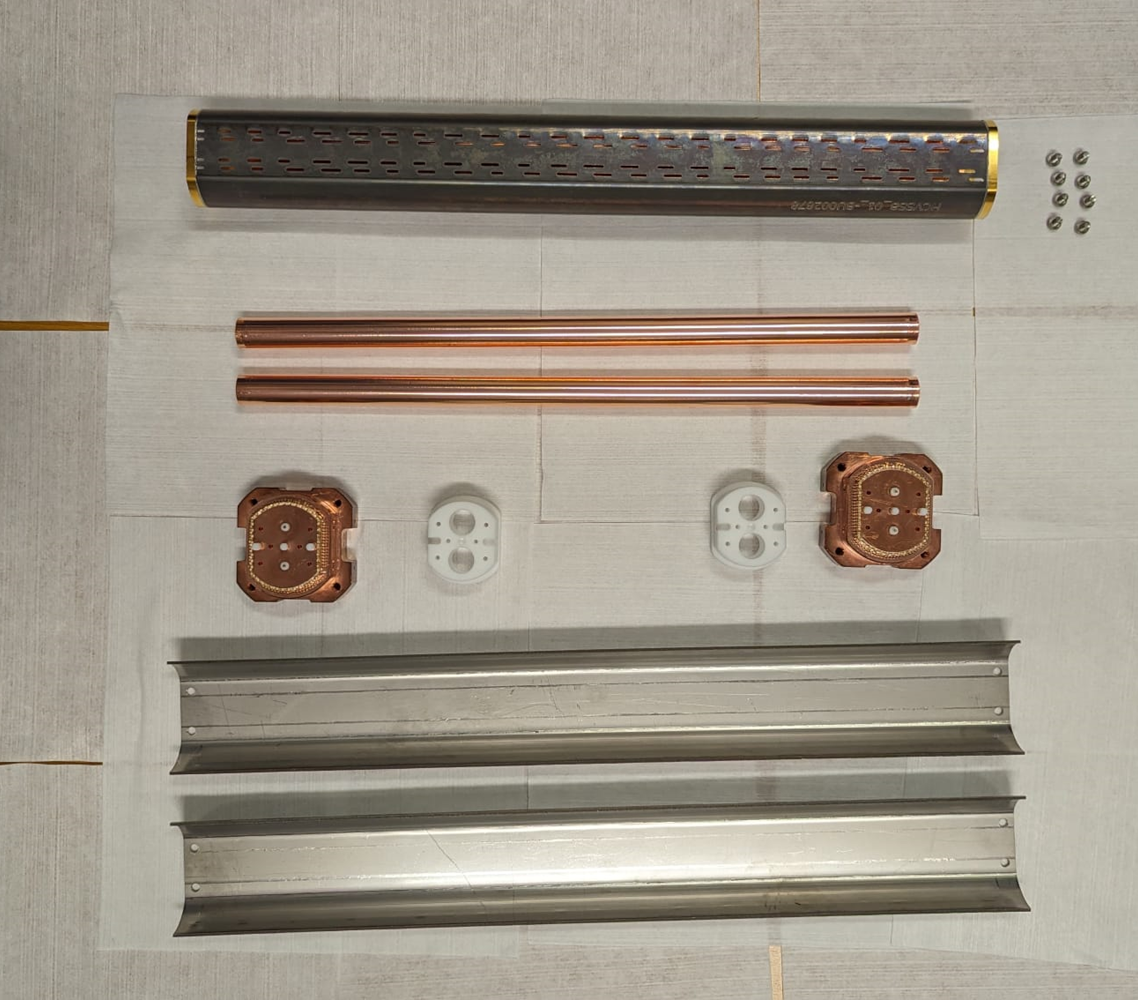}
         \caption{shielded pair assembly pieces}
         \label{fig:assembly1}
     \end{subfigure}
     \hfill
     \begin{subfigure}[b]{0.32\textwidth}
         \centering
         \includegraphics[width=\textwidth]{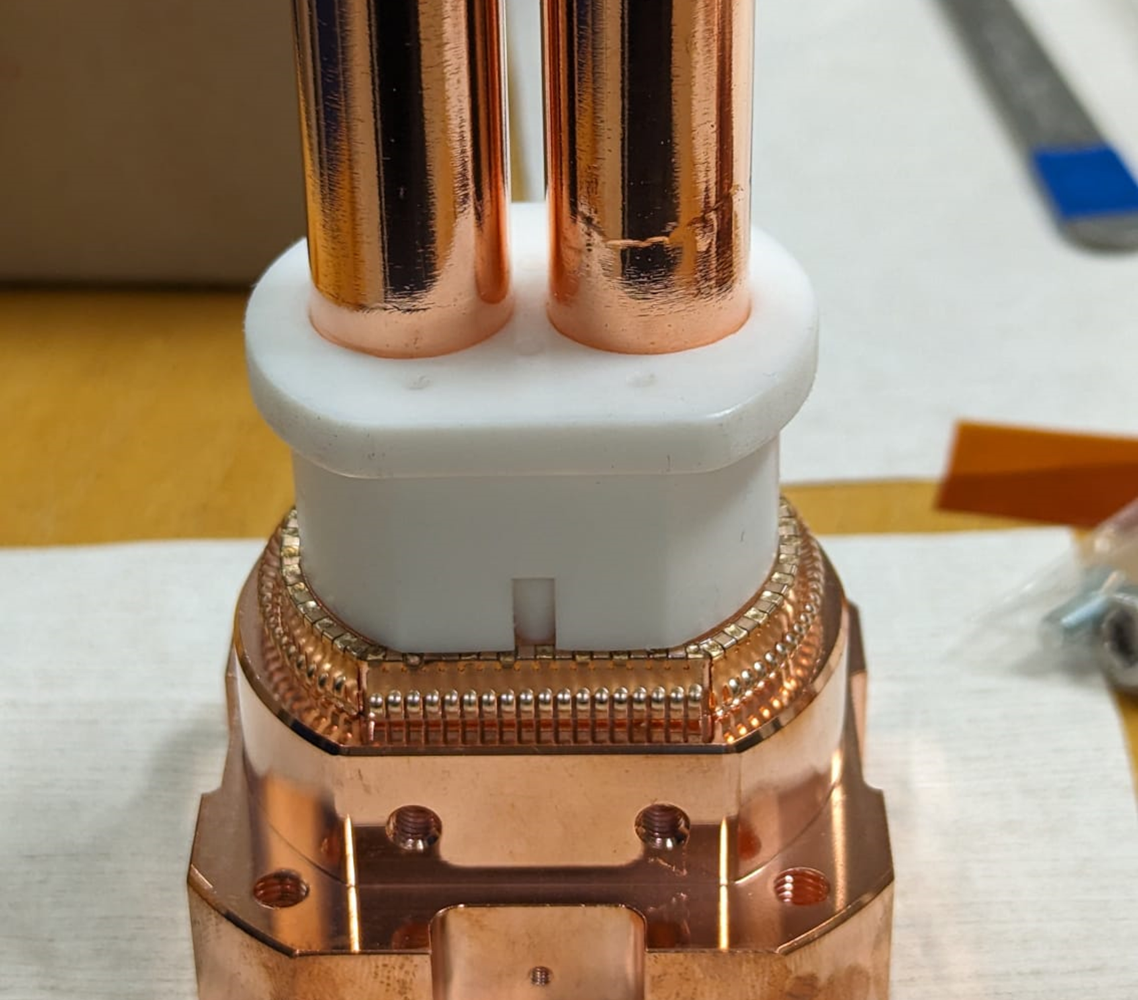}
         \caption{end cap assembly}
         \label{fig:assembly2}
     \end{subfigure}
     \hfill
          \begin{subfigure}[b]{0.32\textwidth}
         \centering
         \includegraphics[width=\textwidth]{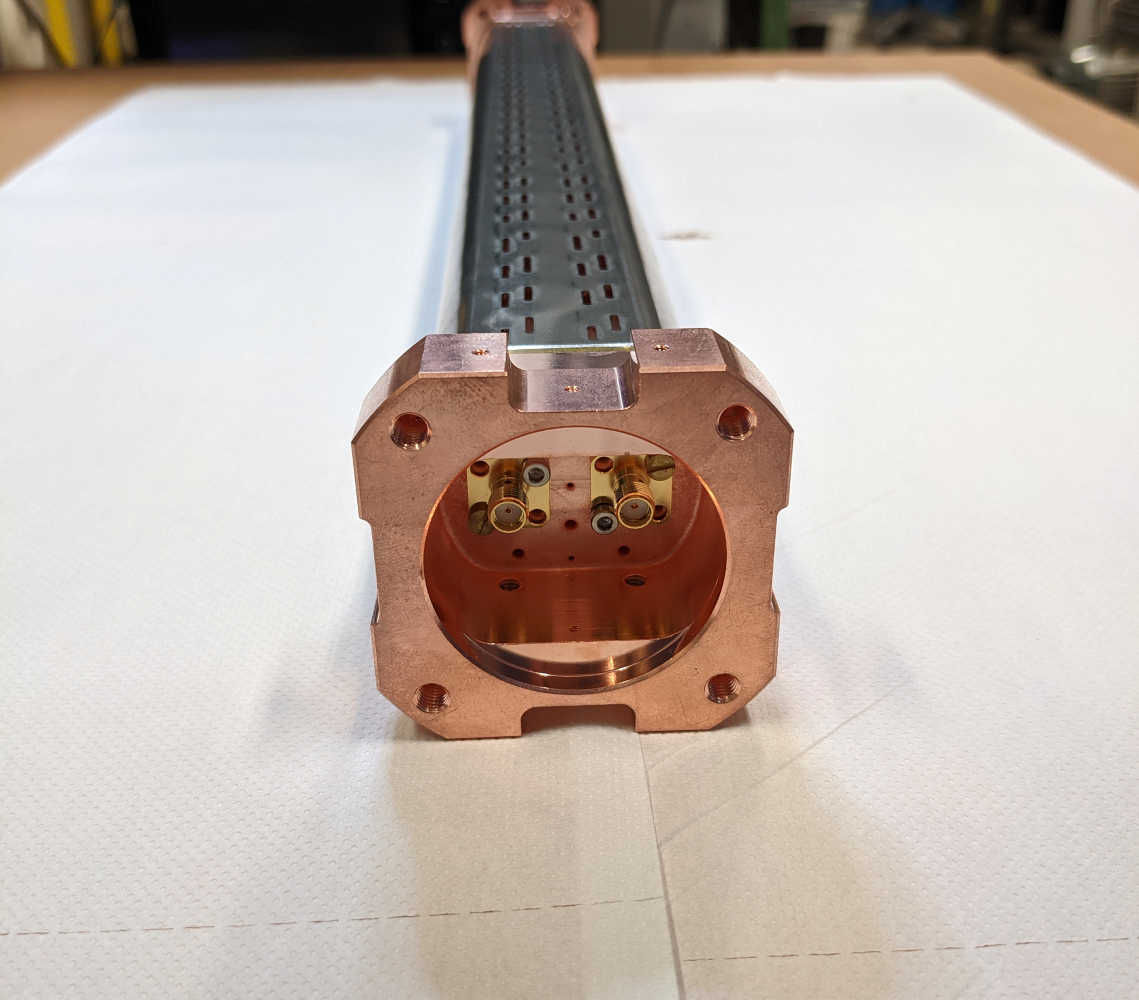}
         \caption{view on probes    }
         \label{fig:assembly3}
     \end{subfigure}
        \caption{Assembly of the LHC beam screen for measurements with the shielded pair technique.}
        \label{fig:assembly}
\end{figure}

\subsection{Experimental Setup}

The shielded pair technique assembly is illustrated in \Cref{fig:assembly}. In this setup, the BS under study serves as the resonator body, enclosed at each end with copper-plated end caps. To ensure electrical continuity, the contact surfaces between the end caps and the BS are equipped with RF fingers. The resonator is excited in specific modes with a detectable resonance frequency $f_0$ and quality factor $Q$ (Q-factor) which allows a rather straightforward calculation of the surface resistance $R_S$. Detailed steps for extracting the surface resistance from these measurements are provided in Brunner \textit{et al.} \cite{KristofMDPI}. 

Two hollow, copper-plated stainless steel rods are inserted along the centre of the BS for two purposes: (1) to adjust resonance frequencies to fall within the relevant LHC beam spectrum (i.e., below 2 GHz), and (2) to allow excitation of two resonances that are closely spaced in frequency. This enables the separation of losses associated with the rods and the independent quantification of losses within the BS. These rods are aligned and secured by two Teflon support elements positioned at either end, as shown in \Cref{fig:assembly2}. The supports are azimuthally locked in place with three Teflon pins to prevent rotation.

For additional mechanical stability and to seal the pumping holes on the top and bottom of the LHC beam screen, two stainless steel half-profiles are attached to these surfaces. Electromagnetic (EM) coupling is facilitated through four SMA probes, with two mounted on each end cap, as depicted in \Cref{fig:assembly3}.

\begin{figure}[t!]
    \centering
    \includegraphics[width=\textwidth]{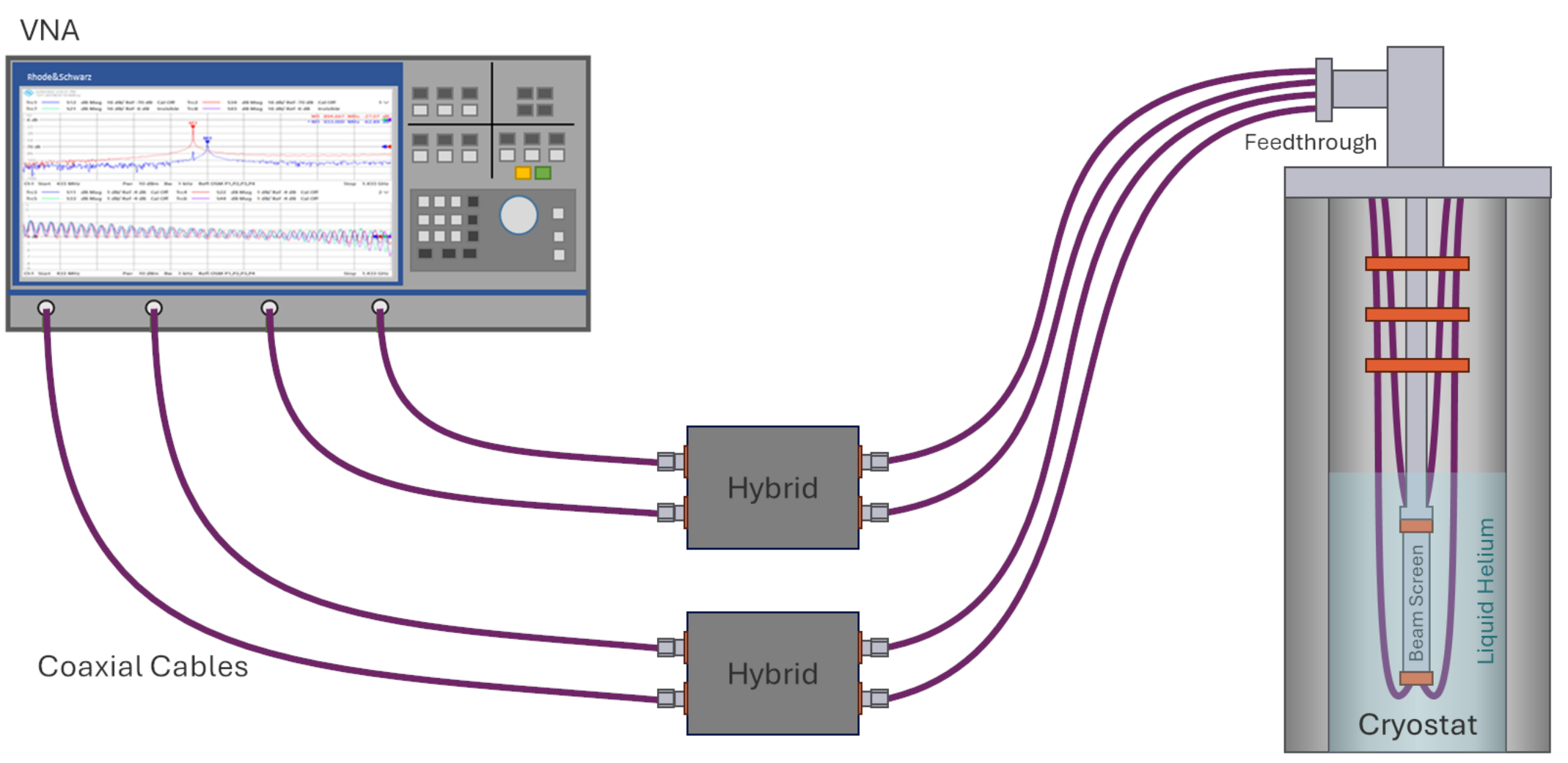}
    \caption{Schematic of the experimental setup.}
    \label{fig:setup}
\end{figure}

\Cref{fig:setup} schematically illustrates the experimental setup, where the BS resonator is connected to a 4-port vector network analyser (VNA) (R\&S®ZNB4, 9\,kHz to 4.5\,GHz) for measurements. Four semi-rigid coaxial cables connect the VNA ports to the input ports of two hybrid devices (Macom H-183-4, 30–3000\,MHz). These hybrid circuits split the input signal into two equal-magnitude outputs, either in-phase ($0^\circ$) or out-of-phase ($180^\circ$). Another set of four coaxial cables connects the hybrids’ output ports to SMA feedthroughs attached to the liquid helium cryostat. Inside the cryostat, four semi-rigid cables connect to the four probes of the BS resonator immersed in liquid helium. The relevant S-parameters are recorded and post-processed using the Algorithm for Resonator Parameter Extraction (ARPE) \cite{Krkotic_ARPE}. This open-source Python-based tool \cite{GitHub}, accessible via a web-based application \cite{ARPE}, was developed to extend the use of microwave resonators beyond the engineering community. Its remote execution capability also prevents code duplication by occasional users unaware of prior work in the field.

\subsection{Numerical Study \label{sec:cst}}

Prior to the experiments, a numerical simulation study was conducted to evaluate the potential effects of the laser treatments. The comparison of numerical data with experimental results will be used to determine the equivalent surface resistances of the Laser-Engineered Surface Structured (LESS) areas, denoted as $R_S^{\text{LESS}}$.

\begin{figure}[t!]
    \centering
    \includegraphics[width=0.9\textwidth]{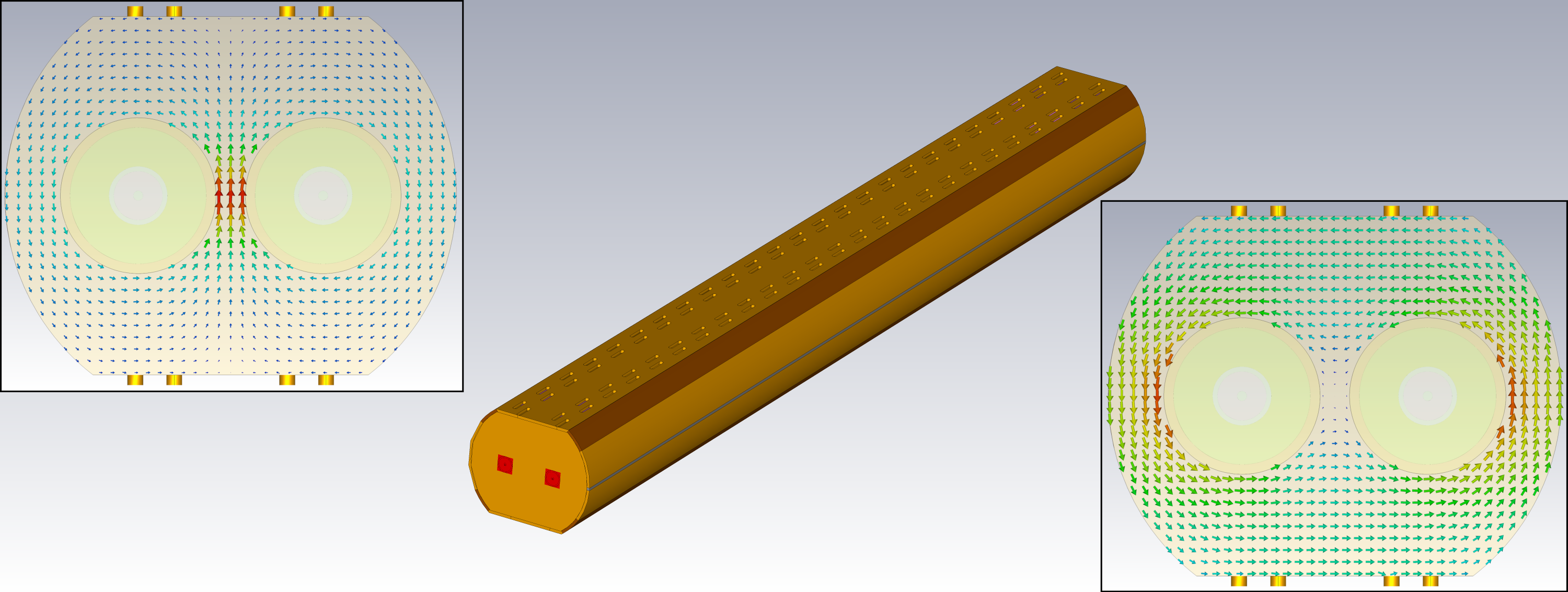}
    \caption{CST version of the reconstructed LHC beam screen with the 4 $\times$ 20$^\circ$ longline treatment. Orange areas represent pure copper, brown areas refer to the laser treated copper and the grey area represents the welding line of stainless steel. Insets indicate the magnetic fields for the even mode (down right) and odd mode (up left) mode configuration.}
    \label{fig:CST}
\end{figure}

\Cref{fig:CST} shows an example of the reconstructed 4 $\times$ 20$^\circ$ longline LHC BS modelled in the 2024 version of the Computer Simulation Tool (CST) Eigenmode Solver. The insets in the figure depict the magnetic field configurations for the two modes of interest. The placement of the rods results in the maximum surface current density being concentrated horizontally on the BS walls.

In the fundamental mode, the vertical distribution of surface current reaches a maximum at the centre, gradually decreasing toward the edges. In the second mode (harmonic), there are two peaks: one on either side of the centre, with a minimum in between. In the third mode, three peaks emerge: one at the centre and two flanking it on either side. This pattern continues with each successive mode introducing an additional peak. It is important to note that this configuration does not replicate the surface current distribution induced by the circulating proton beam.

\begin{figure}[t!]
  \centering
    \includegraphics[width=0.9\textwidth]{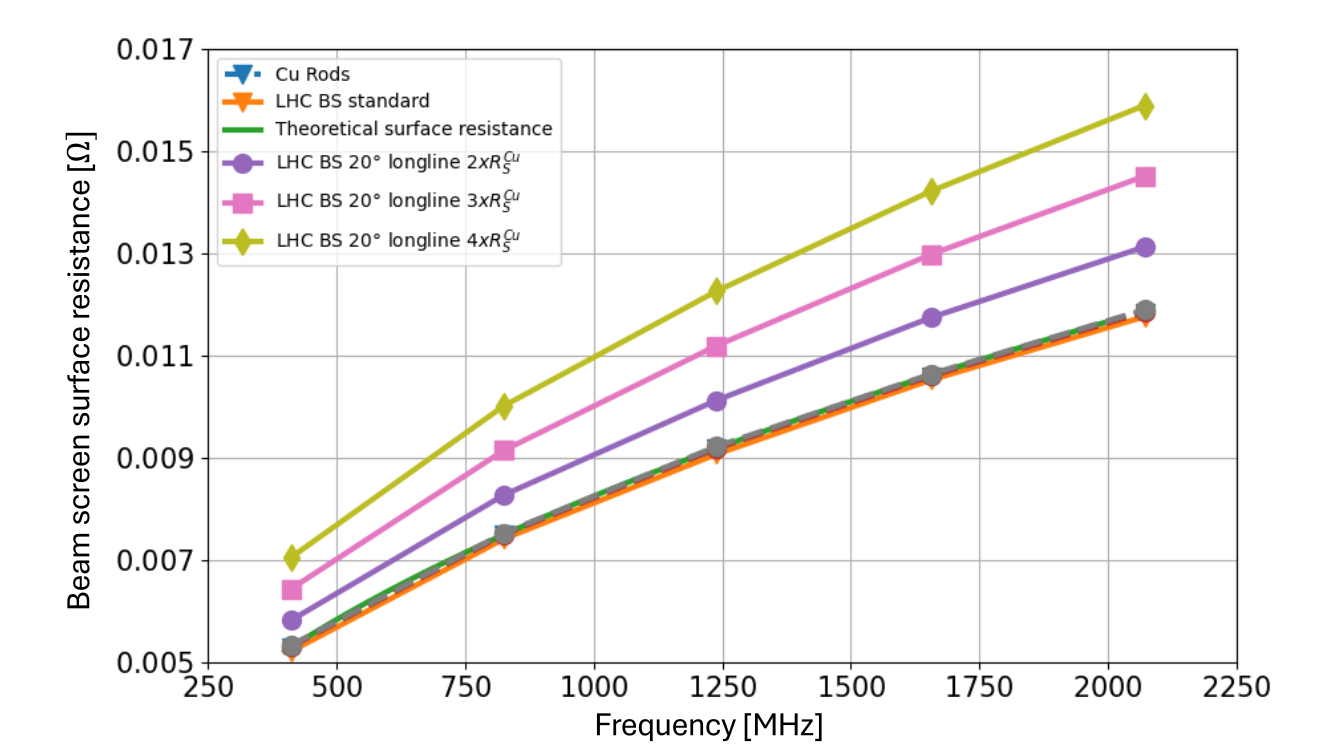}
    \caption{Simulation results for the beam screen surface resistance as a function of effective surface resistance for the 4 $\times$ 20$^\circ$ longline areas at room temperature.}
    \label{fig:CSTResult}
\end{figure}

A study was conducted in which the equivalent $R_S^{\text{LESS}}$ was incrementally increased in integer steps relative to copper's surface resistance $R_S^{\text{Cu}}$. The results for the 4 $\times$ 20$^\circ$ longline treatment at the four corners are illustrated in \Cref{fig:CSTResult}. The figure presents the overall BS surface resistance $R_S^{\text{BS}}$ versus frequency for different increments of $R_S^{\text{LESS}}$ determined using the shielded pair method at RT. The BS surface resistance increases linearly with increase of surface resistance in the laser-treated area. A direct comparison of the BS surface resistances with and without the laser-treatment shall be introduced as follows:
\begin{equation} \label{eq:comp}
   \Delta R_{S}^{\text{BS}} = \frac{R_{S,\text{LESS}}^{\text{BS}} - R_{S,\text{LHC}}^{\text{BS}}}{R_{S,\text{LHC}}^{\text{BS}}}.
\end{equation}
Here, $\Delta R_{S}^{\text{BS}}$ refers to the relative change of BS surface resistance of laser treated BSs $R_{S,\text{LESS}}^{\text{BS}}$ compared to the standard LHC BS $R_{S,\text{LHC}}^{\text{BS}}$. 

\begin{figure}[t!]
     \centering
     \begin{subfigure}[b]{0.49\textwidth}
         \centering
         \includegraphics[width=\textwidth, trim={0 0 1.8cm 0}, clip]{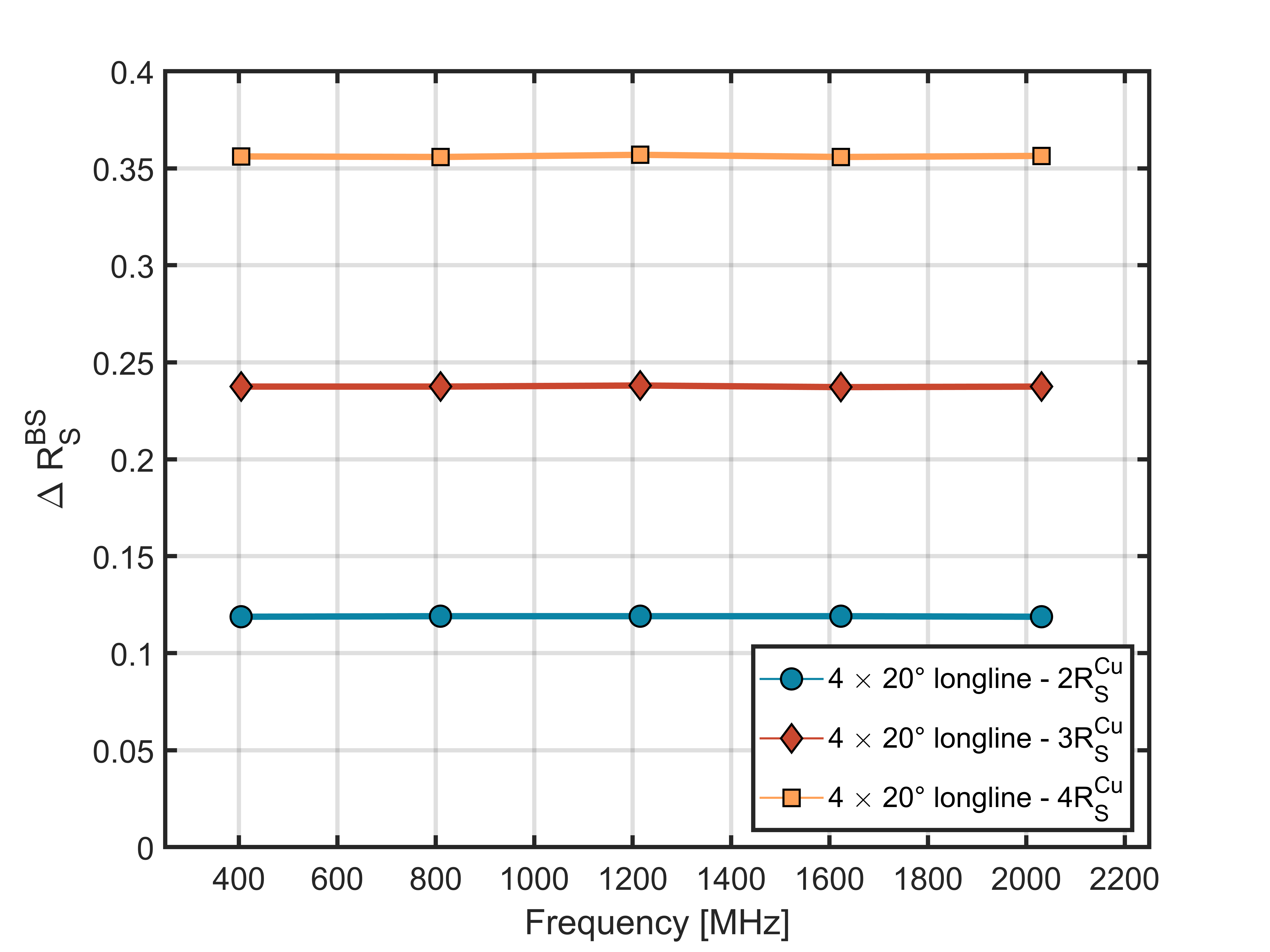}
         \caption{4 $\times$ 20$^\circ$ longline}
         \label{fig:CSTResultsComp1}
     \end{subfigure}
     \hfill
     \begin{subfigure}[b]{0.49\textwidth}
         \centering
         \includegraphics[width=\textwidth, trim={0 0 1.8cm 0}, clip]{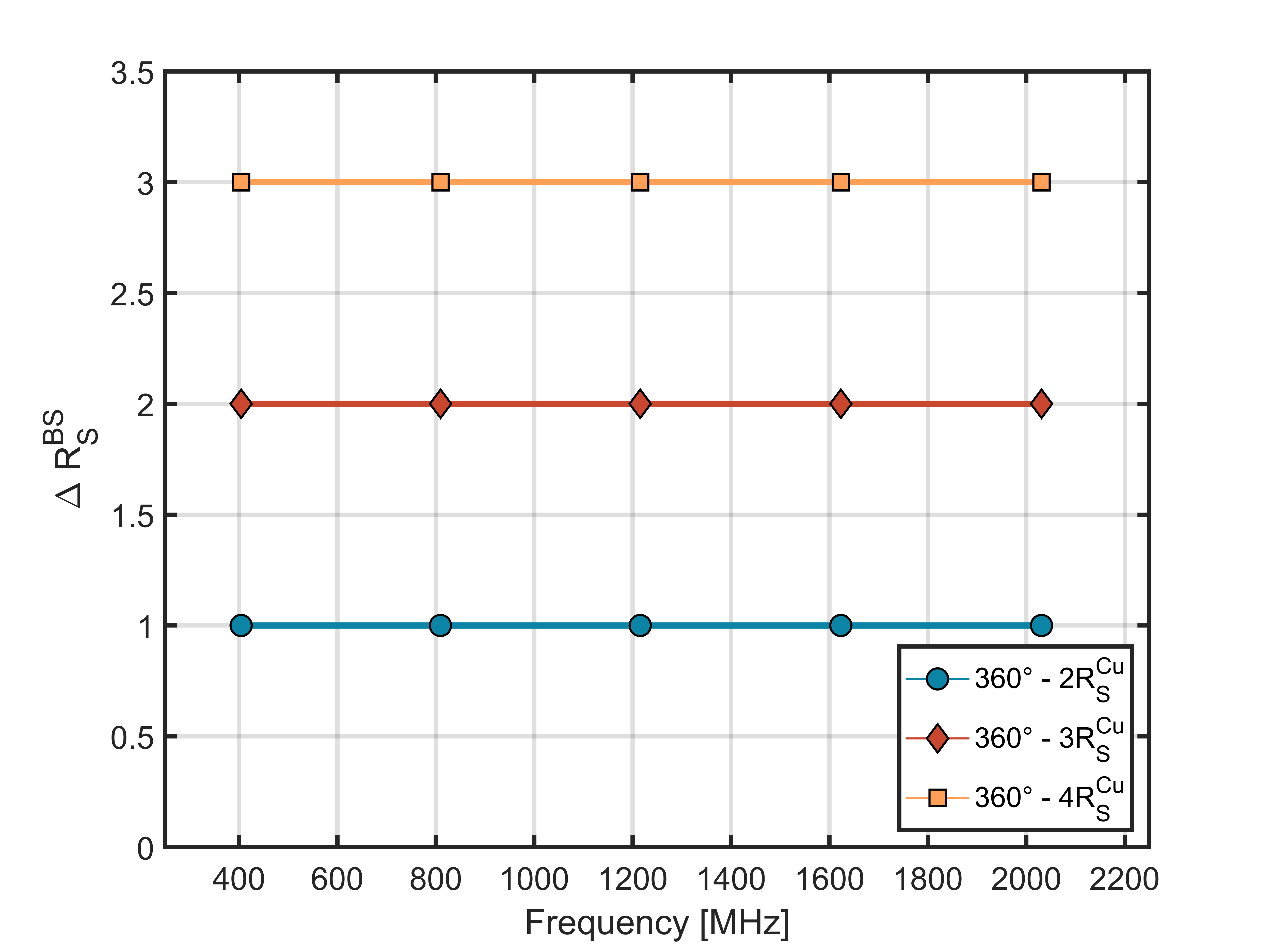}
         \caption{360$^\circ$ longline and spiral}
         \label{fig:CSTResultsComp2}
     \end{subfigure}
        \caption{Direct comparison between the standard and LESS beam screens based on simulations.}
        \label{fig:CSTResultsComp}
\end{figure}

The direct comparison as defined in \Cref{eq:comp} is presented in \Cref{fig:CSTResultsComp}. \Cref{fig:CSTResultsComp1} shows that doubling the surface resistance of the laser-treated copper in the treated zones at the four corners results in approximately a 12\% increase in $R_S^{\text{BS}}$. Tripling the surface resistance yields about a 24\% increase, while a fourfold increase leads to nearly a 36\% rise. Similarly, for both 360° laser treatments as shown in \Cref{fig:CSTResultsComp2}, where the entire BS surface is treated, the simulation indicates that increasing the surface resistance by factors of 2, 3, and 4 results in $R_{S,\text{LESS}}^{\text{BS}}$ increasing by 100\%, 200\%, and 300\%, respectively.

\section{Measurement Results \label{sec:4}}

The measurement procedure for each BS involves five consecutive measurements at RT, followed by cooling down to 4.2\,K and another five measurements with the BS fully submerged in liquid helium to minimise measurement noise. Each set of five measurements is averaged and then post-processed to estimate the $\Delta R_S^{\text{BS}}$ values. For further details on sensitivity and error analysis for this setup, see Brunner \textit{et al.} \cite{KristofMDPI}.

\begin{figure}[b!]
     \centering
     \begin{subfigure}[b]{0.49\textwidth}
         \centering
         \includegraphics[width=\textwidth]{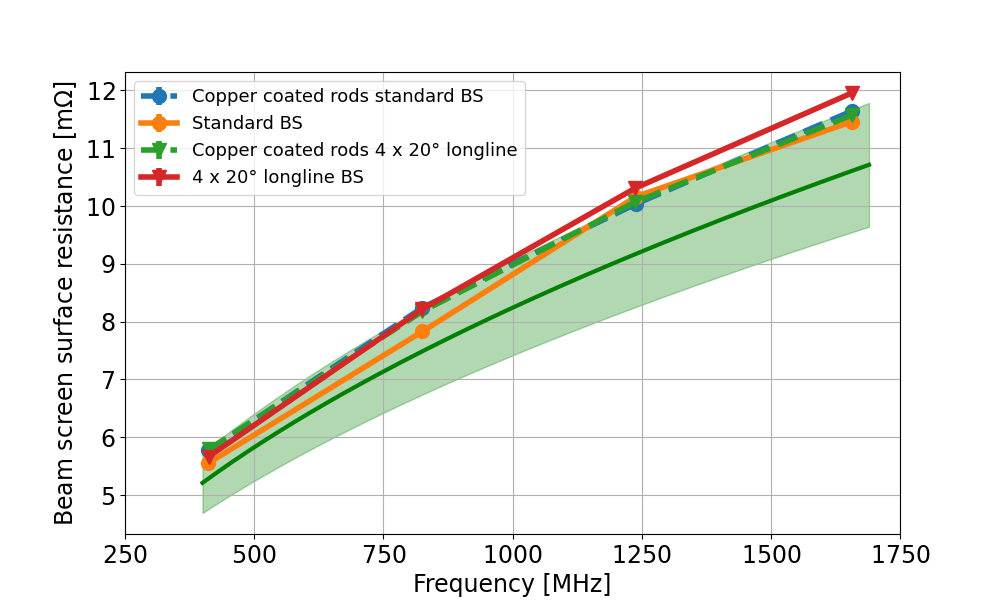}
         \caption{LHC BS and 4 $\times$ 20$^\circ$ longline BS @ RT}
         \label{fig:measresult1}
     \end{subfigure}
     \hfill
     \begin{subfigure}[b]{0.49\textwidth}
         \centering
         \includegraphics[width=\textwidth]{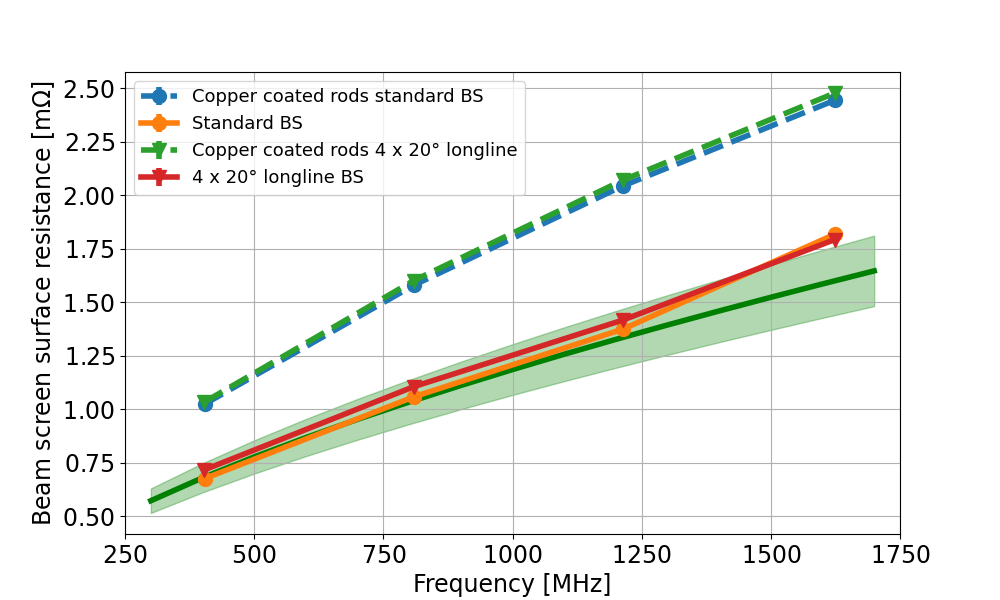}
         \caption{LHC BS and 4 $\times$ 20$^\circ$ longline BS @ 4.2\,K}
         \label{fig:measresult2}
     \end{subfigure}
     \begin{subfigure}[b]{0.49\textwidth}
         \centering
         \includegraphics[width=\textwidth]{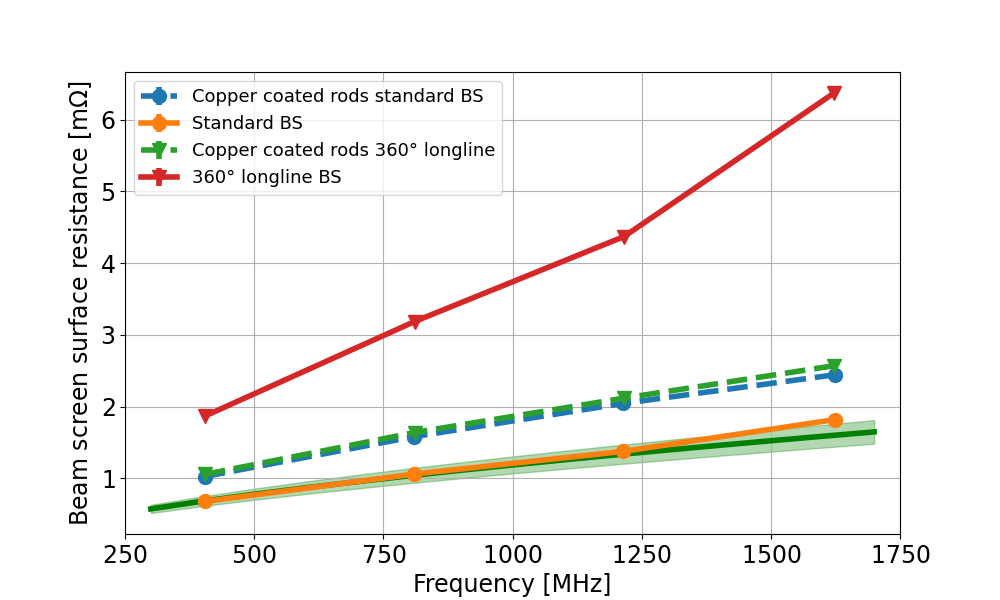}
         \caption{LHC BS and 360$^\circ$ longline BS @ 4.2\,K}
         \label{fig:measresult3}
     \end{subfigure}
     \begin{subfigure}[b]{0.49\textwidth}
         \centering
         \includegraphics[width=\textwidth]{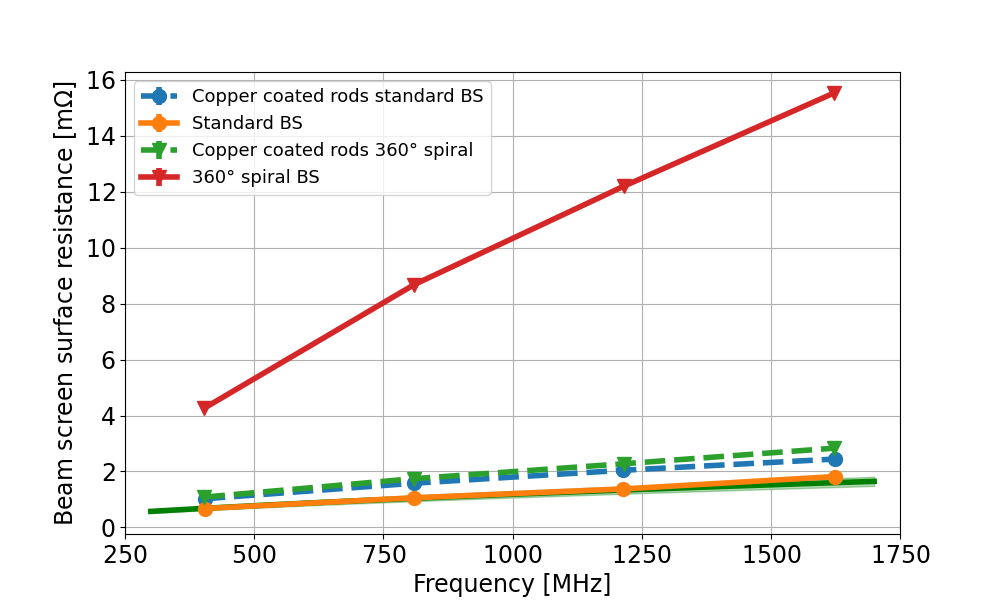}
         \caption{LHC BS and 360$^\circ$ spiral BS @ 4.2\,K}
         \label{fig:measresult4}
     \end{subfigure}
        \caption{Measurement results, showing a direct comparison between each laser-treated beam screen and the standard beam screen.}
        \label{fig:measresult}
\end{figure}

\Cref{fig:measresult} provides an overview of the measurement results. The figure shows the measured surface resistance of the laser-treated BS samples (red curves) in comparison with the standard BS (orange curves). The surface resistance of the copper-coated rods, additionally named by the corresponding BS sample in the legend, is represented by dashed lines. For reference, the theoretically expected surface resistance of pure copper is indicated by the green solid line with an error band of expectable values.

\Cref{fig:measresult1} shows a representative example of the BS surface resistance measurement for the standard LHC BS and 4 $\times$ 20$^\circ$ longline treatment at RT. \Cref{fig:measresult2}, \Cref{fig:measresult3} and \Cref{fig:measresult4} show the measured results at 4.2\,K. Here, the longitudinal weld in the midplane of the BS of 2.1\,mm width is calculated out \cite{KristofMDPI}. 

In \Cref{fig:measresult2}, the data indicates that the 4 $\times$ 20$^\circ$ longline BS exhibits a slightly higher surface resistance than the untreated BS at both temperatures. The results for the copper coated rods show consistent results between the measurements. \Cref{fig:measresult3} and \Cref{fig:measresult4} show the strong impact of the laser treatments on the surface resistance.

\begin{figure}[h!]
     \centering
          \begin{subfigure}[b]{0.49\textwidth}
         \centering
         \includegraphics[width=\textwidth]{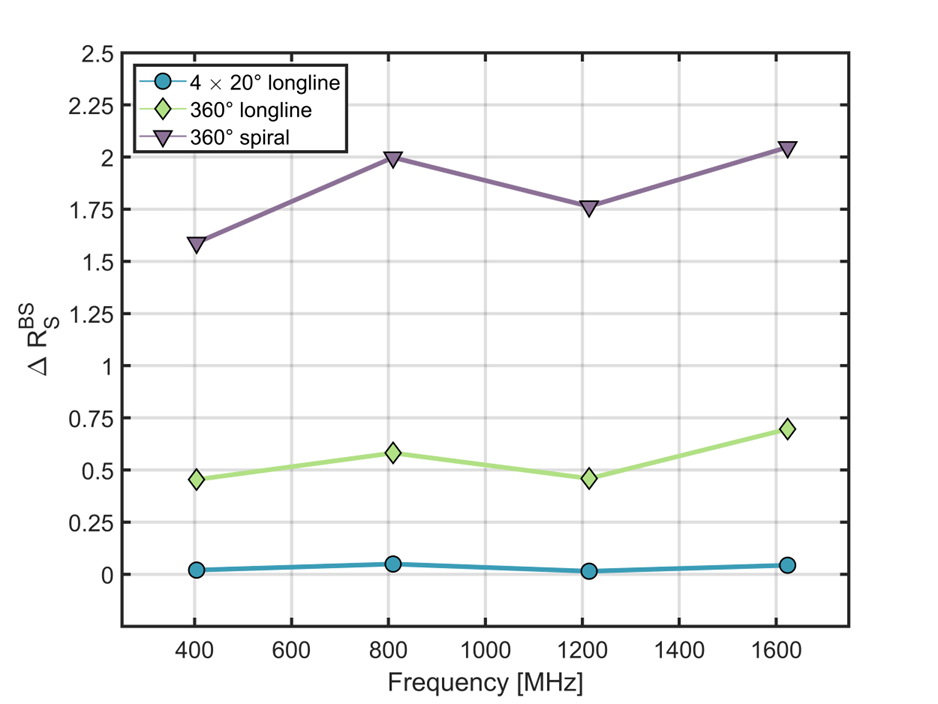}
         \caption{direct comparison @ RT}
         \label{fig:measresultc1}
     \end{subfigure}
     \hfill
      \begin{subfigure}[b]{0.49\textwidth}
         \centering
         \includegraphics[width=\textwidth]{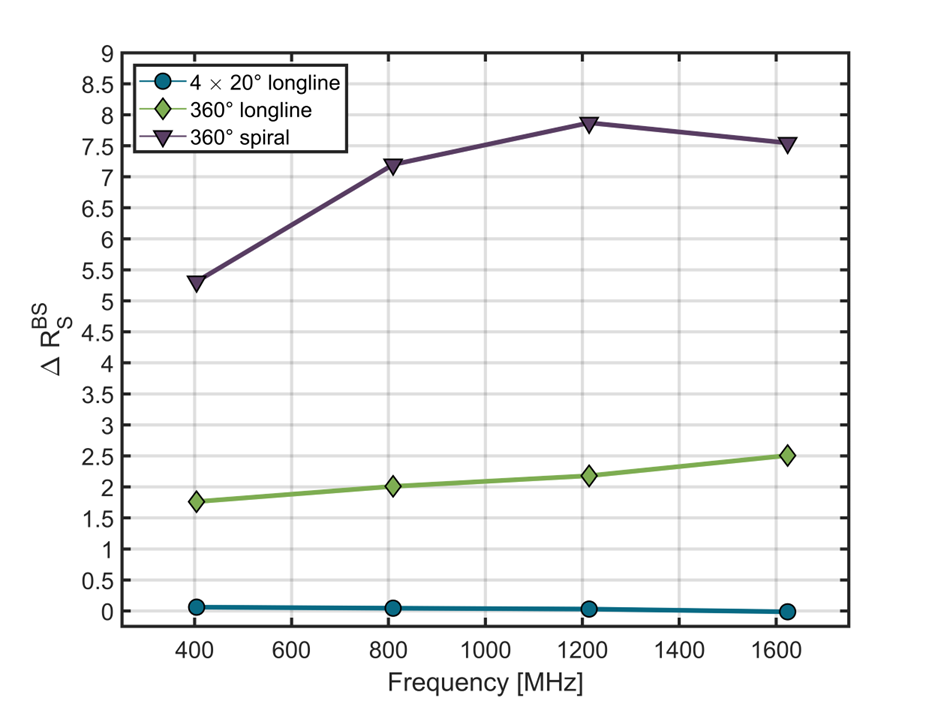}
         \caption{direct comparison @ 4.2\,K}
         \label{fig:measresult2c}
     \end{subfigure}
        \caption{Measurement results presented as relative change of BS surface resistance for the laser treated samples compared to the standard LHC BS.}
        \label{fig:measresultc}
\end{figure}

To facilitate the comparison \Cref{fig:measresultc1} and \Cref{fig:measresult2c} present the relative surface resistance  $\Delta R_{S}^{\text{BS}}$, as defined in \Cref{eq:comp}, at RT and 4.2\,K, respectively. The results reveal a modest average increase of 3\,\% when treating the four corners of the BS at both temperatures. The 360$^\circ$ longline treatment exhibits intermediate results, with an average increase of 55\,\% (+14\,\% | -9\,\%) at RT and 211\,\% (+39\,\% | -35\,\%) at 4.2\,K. By contrast, the 360$^\circ$ spiral treatment leads to an average increase of 185\,\% (+19\,\% | -26\,\%) at RT and 698\,\% (+89\,\% | -167\,\%) at cryogenic temperature. 

The corresponding increase in surface resistance needed for the laser-treated areas to match these results has been determined using the numerical estimations from \Cref{sec:cst}. These findings are summarised in \Cref{tab:results}, which shows the factor by which the surface resistance increases relative to pure copper, i.e, for example, $R_{S,\text{4.2\,K}}^{\text{360° spiral}} = 7.98 \times R_{S,\text{4.2\,K}}^{\text{Cu}}$.

\setlength{\tabcolsep}{12pt}
\begin{table}[h!]
    \centering
    \caption{Summary of the surface resistance increase factor for the laser-treated areas, relative to pure copper, averaged across all resonances. The error is defined as the maximum deviation from the average to the highest or lowest value.}
    \label{tab:results}
    \begin{tabular}{cccc}
    \toprule \toprule
         & \textbf{4 $\times$ 20$^\circ$ longline}  & \textbf{360$^\circ$ longline} & \textbf{360$^\circ$ spiral} \\ \midrule 
       $R_{S,\text{RT}}^{\text{LESS}}$  & $1.29\substack{+0.16 \\ -0.16}$ & $1.55\substack{+0.14 \\ -0.09}$ & $2.85\substack{+0.19 \\ -0.26}$ \\
       $R_{S,\text{4.2\,K}}^{\text{LESS}}$  & $1.41\substack{+0.15 \\ -0.13}$ & $3.11\substack{+0.39 \\ -0.35}$  & $7.98\substack{+0.89 \\ -1.67}$\\ \bottomrule \bottomrule
    \end{tabular}
\end{table}

\section{Discussion \label{sec:5}}
The findings of the study demonstrated that for laser-treated surfaces the extent of the treated area, the groove depth, and particularly the groove orientation relative to the current flow have a significant impact on the equivalent surface resistance. For instance, the 4 $\times$ 20$^\circ$ longline laser treatment, characterised by the smallest groove depths and treatment area, exhibited the most favourable equivalent surface resistance. In contrast, the 360$^\circ$ spiral laser treatment exhibited the highest equivalent surface resistance, with a particularly large difference when compared to untreated copper, especially at cryogenic temperatures. Notably, at cryogenic temperatures, this sample's surface resistance approached the level of pure copper at room temperature. 

It is important to emphasise that the grooves are aligned perpendicular to the induced current during measurements. By merely rotating the groove directions, as in the 360$^\circ$ longline laser treatment, the equivalent surface resistance produced intermediate results due to the parallel alignment with the current. Even though, the ranking of the results presented in the previous section aligns with expectations based on findings from previous studies \cite{SergioQPR, Krkotic_PRAB} and shall be further discussed in the following section, additional factors related to the measurement technique influenced this ranking. 

Madarasz \textit{et al.} \cite{Tamas} observed that misalignment between grooves and the surface current density, resulting in intersections between the current and grooves, increases the measured surface resistance. Similarly, the results for the 360$^\circ$ longline laser treatment may be affected by the alignment precision between the inner rods and the laser treatment, as achieving perfect parallelism between the surface current and the grooves is inherently challenging.

\begin{figure}[p!]
     \centering
     \begin{subfigure}[b]{\textwidth}
         \centering
         \includegraphics[width=\textwidth, trim = {0 6cm 0 6cm}, clip ]{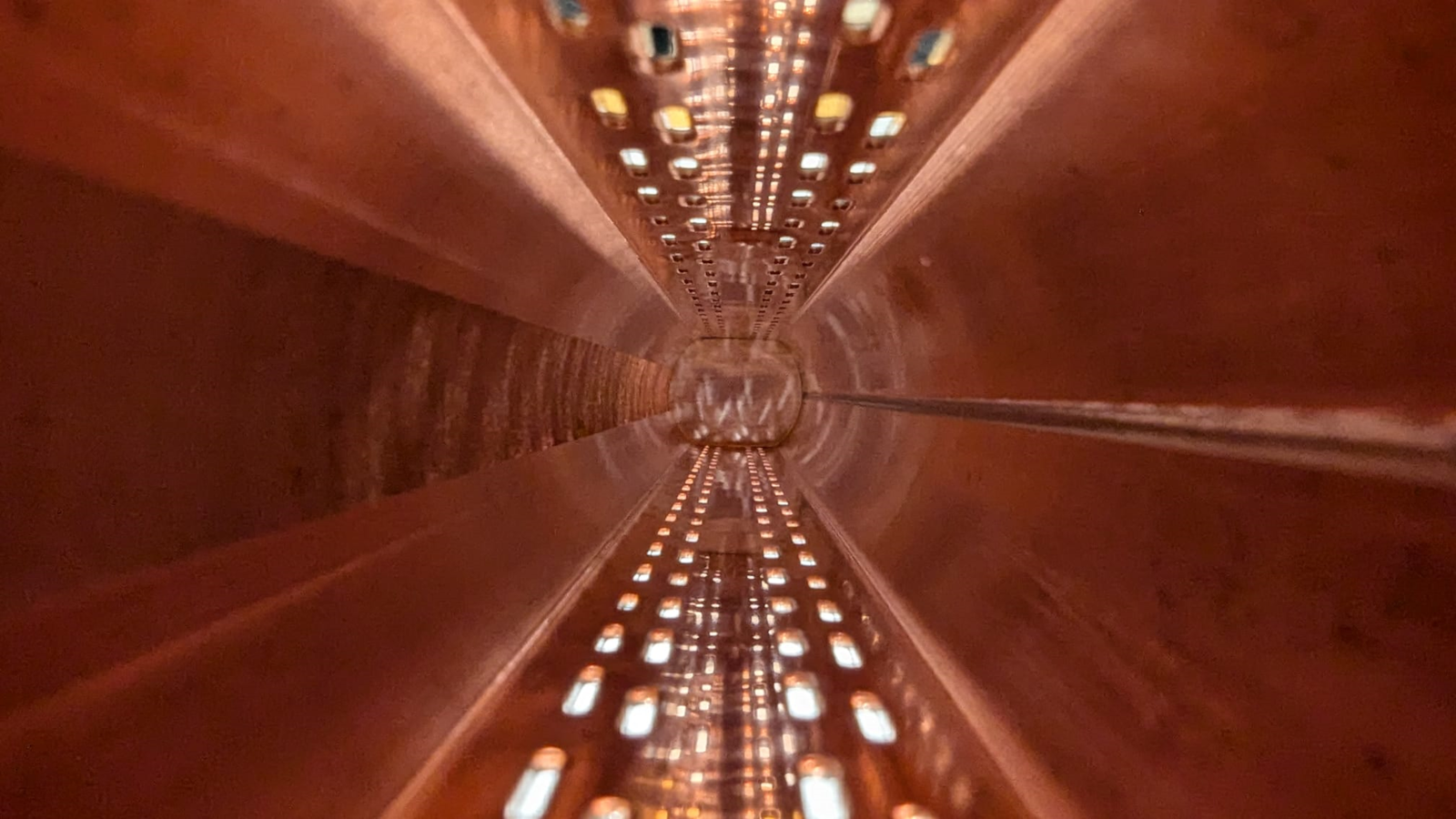}
         \caption{standard LHC beam screen}
         \label{fig:micro0}
     \end{subfigure}
     \begin{subfigure}[b]{0.46\textwidth}
         \centering
         \includegraphics[width=\textwidth]{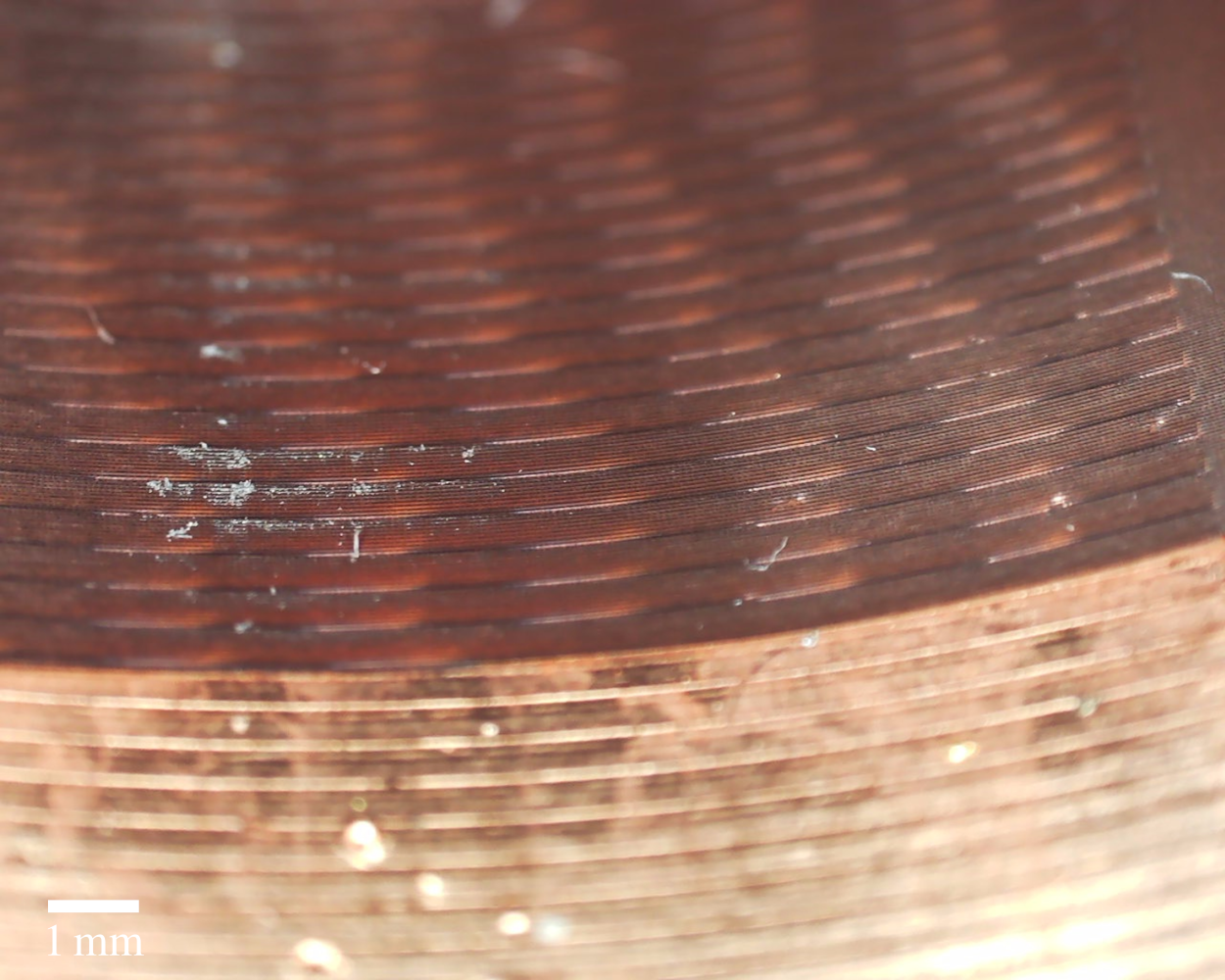}
         \caption{spiral laser treatment of sawtooth area}
         \label{fig:micro1}
     \end{subfigure}
     \hfill
     \begin{subfigure}[b]{0.49\textwidth}
         \centering
         \includegraphics[width=\textwidth]{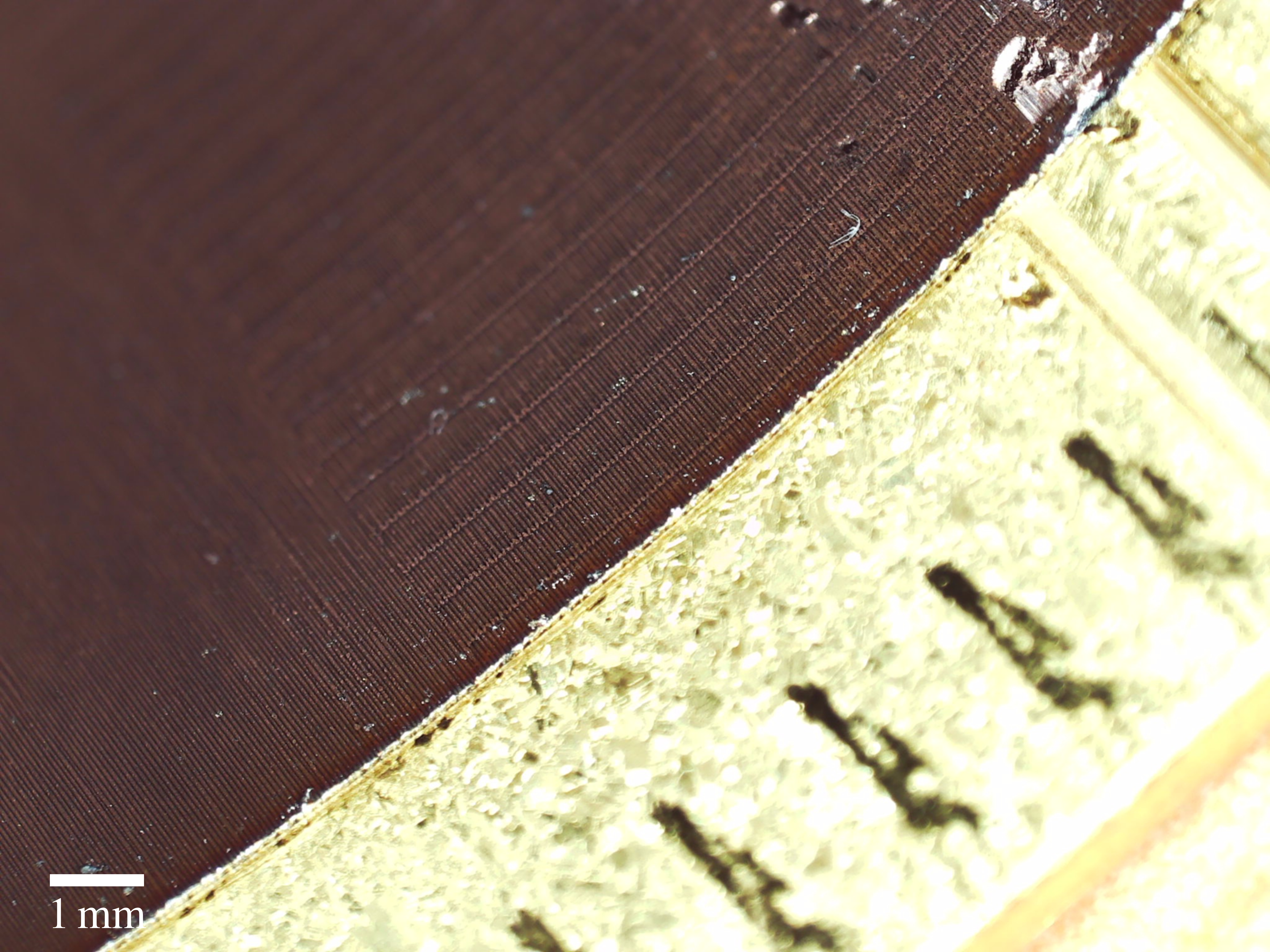}
         \caption{longline laser treatment of sawtooth area}
         \label{fig:micro2}
     \end{subfigure}
     \hfill
          \begin{subfigure}[b]{0.46\textwidth}
         \centering
         \includegraphics[width=\textwidth]{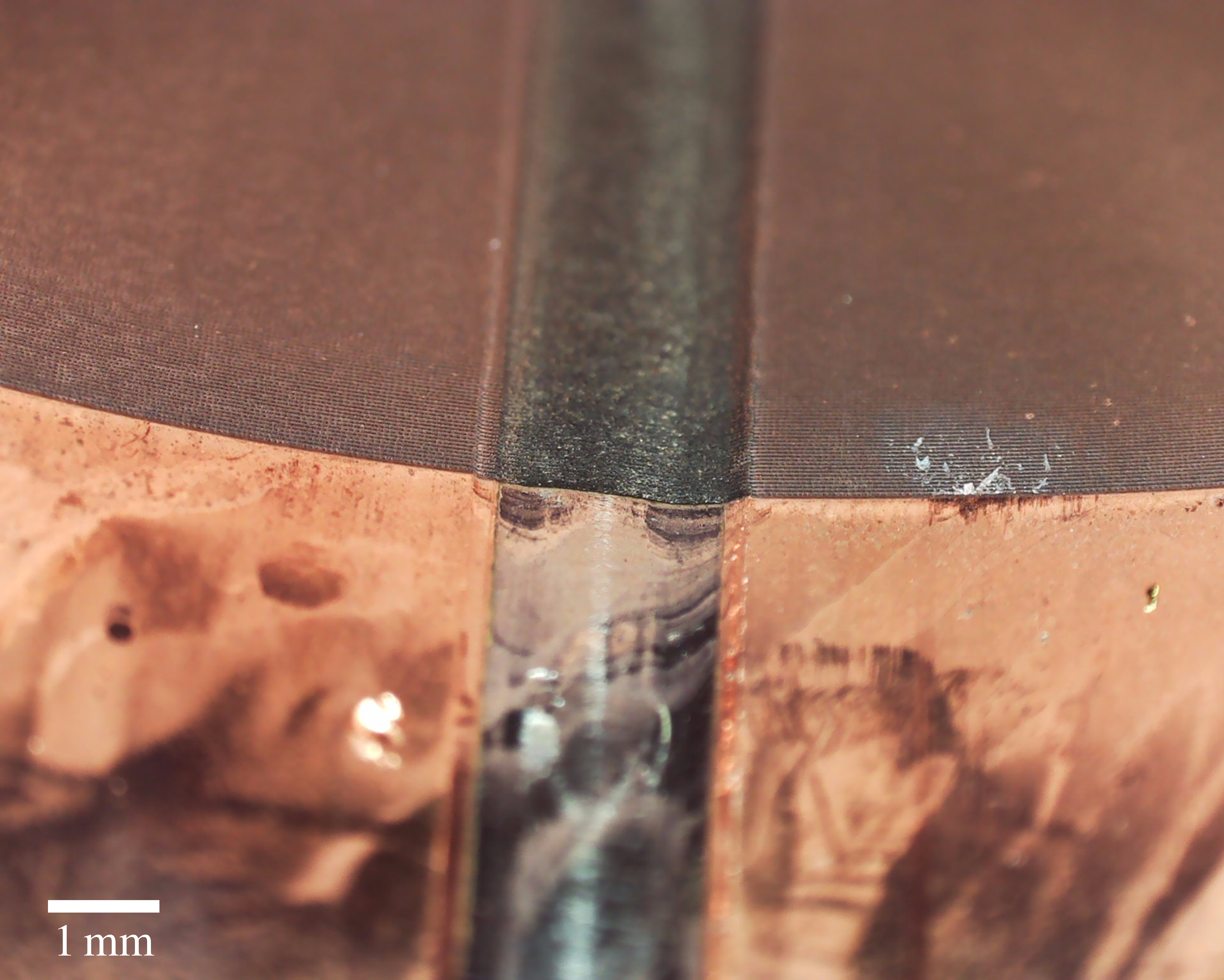}
         \caption{spiral laser treatment of stainless steel weld}
         \label{fig:micro3}
     \end{subfigure}
     \hfill
      \begin{subfigure}[b]{0.49\textwidth}
         \centering
         \includegraphics[width=\textwidth]{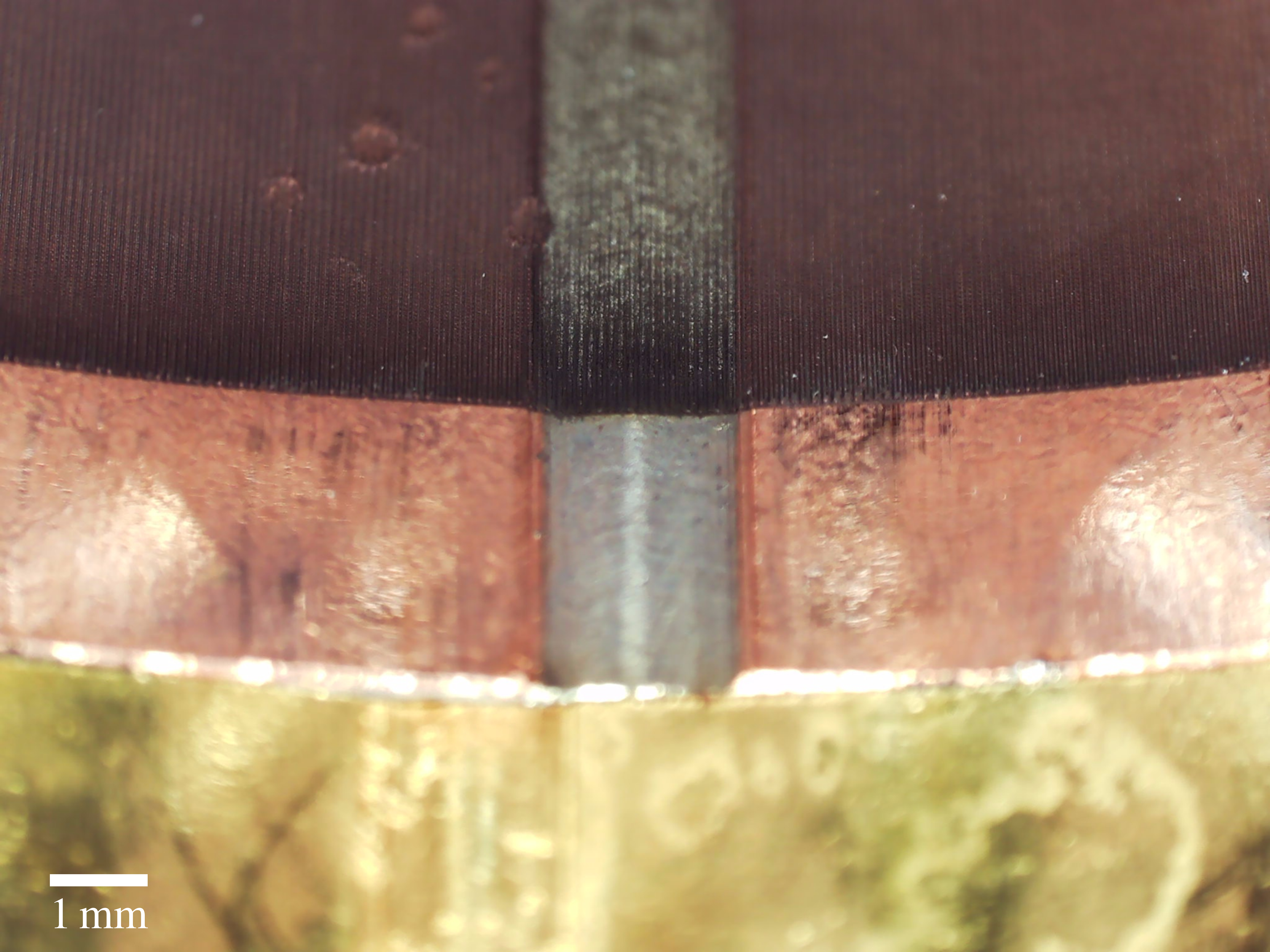}
         \caption{longline laser treatment of stainless steel weld}
         \label{fig:micro4}
     \end{subfigure}
        \caption{Photographs of beam screen samples, highlighting the laser-treated areas and welds in detail.}
        \label{fig:micro}
\end{figure}

In addition to the aforementioned alignment, the two 360$^\circ$ BSs also underwent laser treatment on the sawtooth area and stainless steel welds, as shown in \Cref{fig:micro}. This effect on the measured surface resistance cannot be excluded from the results and very likely increases the measured values above the real values, though the exact extent remains unclear. Separate measurements focusing exclusively on laser treatment on these areas would be required to accurately quantify their influence. 

The sawtooth pattern consists of a longitudinal period of about 500\,\textmu m, a horizontal amplitude of around 35\,\textmu m, and a vertical extent of 7.5\,mm from the equatorial plane \cite{Guillermo, Zimmermann}. Given that the laser treatment penetrates 38 to 67\,\textmu m deep for the two fully treated BSs, it is possible that, at certain locations of the sawtooth pattern, the laser may have penetrated through the copper layer due to the gradual reduction of copper at each sawtooth. This could expose the underlying stainless steel, potentially further increasing the surface resistance when compared to data from other studies. Additionally, the impact of the laser treatment on areas with an inclined angle, such as the sawtooth pattern, has not yet been studied. Micrographs of the two BSs can be seen in \Cref{fig:micro1} and \Cref{fig:micro2} for the 360$^\circ$ spiral and 360$^\circ$ longline, respectively.

On the opposite side of the BS, the longitudinal stainless steel weld in the midplane, with a width of 2.1\,mm \cite{KristofMDPI}, poses a challenge for accurately determining the surface resistance of the laser-treated copper area. As shown in \Cref{fig:micro3} and \Cref{fig:micro4}, the welds are also laser-treated, and their surface resistance remains unknown. Post-processing calculations are based on values determined by Brunner\cite{KristofPhD} for this weld without laser treatment, with the assumption that the surface resistance is constant. However, for the two BSs treated with 360$^\circ$ laser patterns, this assumption only partially accounts for the weld's actual contribution to the overall surface resistance.

\section{Comparison to earlier experiments \label{sec:6}}

In Calatroni \textit{et al.} \cite{SergioQPR}, the validation of laser-treated OFE copper was performed using a superconducting quadrupole resonator (QPR) \cite{Mahner, Junginger} operating at frequencies from 400\,MHz to 1200\,MHz and temperatures ranging from 2\,K to 15\,K. This study overlaps in the cryogenic temperature and frequency with the investigation conducted here. Additionally, Krkoti\'{c} \textit{et al.} \cite{Krkotic_PRAB} performed a study using a sapphire-loaded Hakki-Coleman dielectric resonator (DR) operating at 3.4\,GHz \cite{KristofPRAB}, across a temperature range from room temperature to liquid nitrogen temperature. This study overlaps with the room temperature range of the investigation conducted here, although at a higher frequency. In both studies, two solid OFE copper discs were laser-structured with different patterns: one with a radial pattern (forming triangular sectors) and the other with an azimuthal pattern (single spiral starting from the centre extending outward). Based on the measurement techniques in both arrangements, for radially treated discs the RF currents intersect orthogonal to the laser-etched grooves while for the spirally treated discs, the RF surface currents are aligned with the groove lines, which is the opposite case in the study conducted here.

\begin{figure}[b!]
     \begin{subfigure}[b]{\columnwidth}
         \centering
         \includegraphics[width=\textwidth]{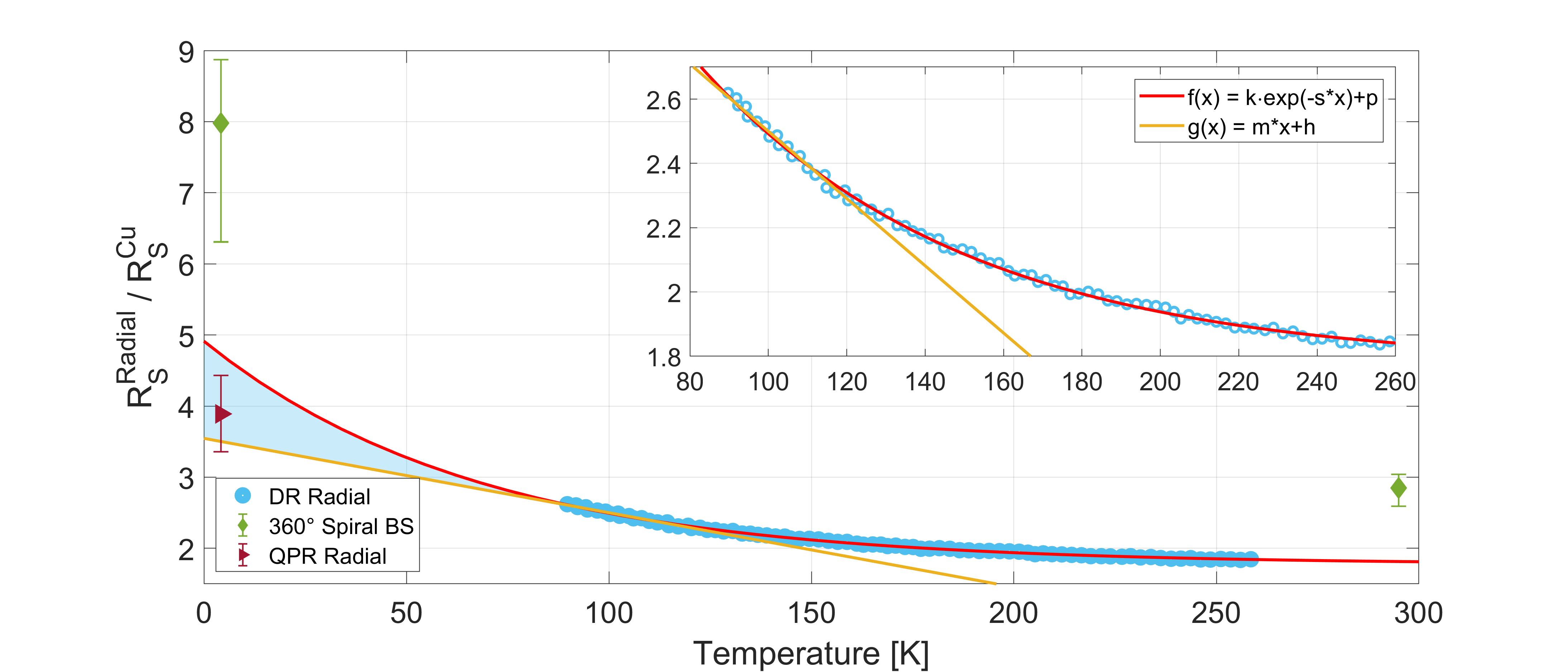}
         \caption{grooves perpendicular to the surface current}
         \label{fig:DRratio2}
     \end{subfigure}
          \centering
     \begin{subfigure}[b]{\columnwidth}
         \centering
         \includegraphics[width=\textwidth]{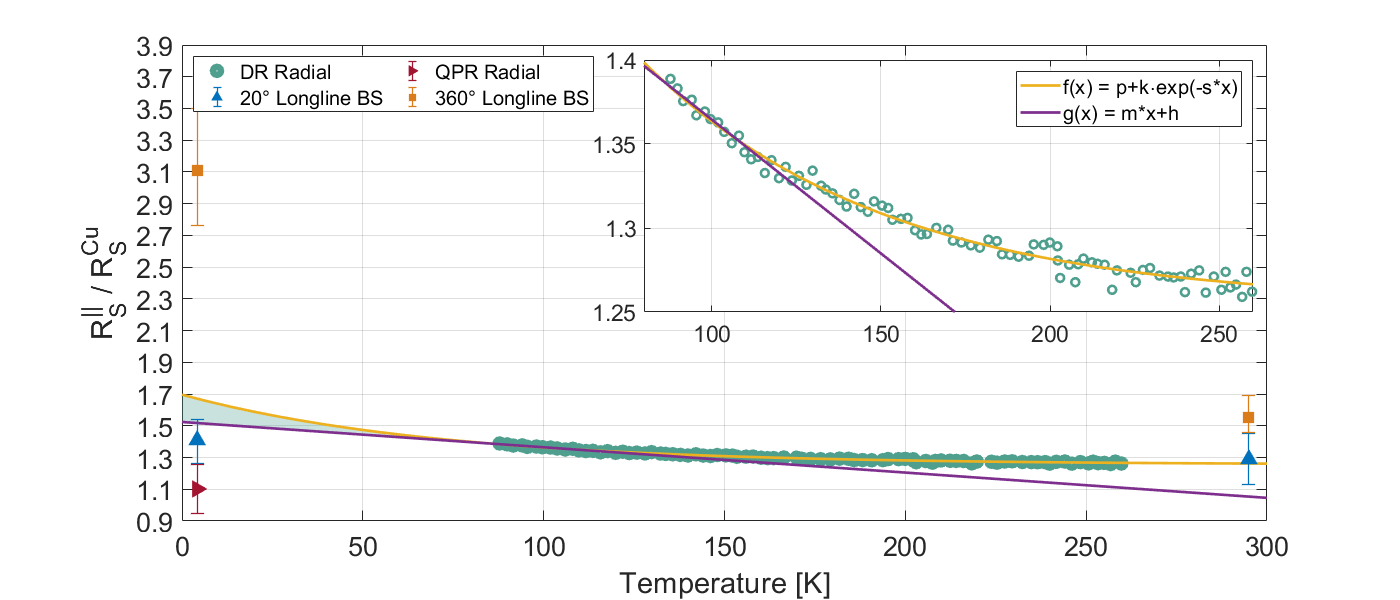}
         \caption{grooves parallel to the surface current}
         \label{fig:DRratio1}
     \end{subfigure}
        \caption{Measurement results obtained using different techniques, including a fitting procedure for predicting surface resistance values.}
        \label{fig:DRratio}
\end{figure}

\Cref{fig:DRratio} contrasts the results from the three studies by presenting the direct ratio of the equivalent surface resistance of laser-treated copper to that of untreated copper, separated by groove orientation relative to the surface current: \Cref{fig:DRratio2} shows the ratio for samples with grooves perpendicular to the surface current, while \Cref{fig:DRratio1} shows the ratio for samples with grooves aligned with the surface current. Additionally, two functions, as indicated in the legend of the inset, were used to fit the data obtained with the DR technique to predict values at cryogenic temperatures for direct comparison. The exponential function $f(x)$ was fitted over the entire temperature range, while the linear function $g(x)$ was fitted to the data obtained between 80\,K and 125\,K. This approach provides an upper and lower rough estimation of the surface resistance values as temperatures approach absolute zero, making it easier to compare the results across studies and assess the consistency of the data. 

In \Cref{fig:DRratio2}, the fitting results for the DR-determined data reveal a potential range between 3.5 to 5 at 4.2\,K. The QPR-measured data also falls within this predicted range. Examining the laser parameters listed in \Cref{tab:laser1}, it can be seen that the same pulse length, wavelength, and repetition rate were used for both sample types. However, the scanning speed for the QPR sample was slower compared to the DR sample, which results in deeper grooves and, consequently, a higher surface resistance compared to the DR sample could be expected. 

\setlength{\tabcolsep}{6pt}
\begin{table}[h!]
    \centering
    \caption{Comparison of laser parameters utilised for the creation of LESS samples.}
    \label{tab:laser1}
    \begin{tabular}{cccccc}
    \toprule \toprule
                                & \textbf{pulse length [ps]}  & \textbf{wavelength [nm]} &  \textbf{repetition rate [kHz]} & \textbf{scanning speed [\nicefrac{mm}{s}]}  \\ \midrule 
QPR \cite{SergioQPR}      &   10                        &     532                  &  200                            & 10\\
DR \cite{Krkotic_PRAB}         &       10                    &      532                 &     200                         &   15\\ 
BS                  &           1                 &     1030                 &              500                &   20 \\
       \bottomrule \bottomrule
    \end{tabular}
\end{table}
\setlength{\tabcolsep}{8pt}
\begin{table}[h!]
\addtocounter{table}{-1}
\renewcommand{\thetable}{\Roman{table}b}
    \centering
    \label{tab:laser2}
    \begin{tabular}{ccccc}
    \toprule \toprule
                                    & \textbf{spot diameter [µm]} & \textbf{power [W]} & \textbf{groove spacing [µm]} & \textbf{groove depth [µm]} \\ \midrule 
QPR \cite{SergioQPR}           &  12                         &      0.4 \nicefrac{TW}{cm$^2$}              &        24                    &    35                      \\
DR \cite{Krkotic_PRAB}            &     52                      &   4                &        45                    &         15-25              \\
BS                       &        55                   &         6.5        &            50                &          17 - 67   \\
       \bottomrule \bottomrule
    \end{tabular}
\end{table}

Another factor to consider is the frequency at which the two techniques operate. Frequency influences the penetration depth of EM fields, known as the skin depth $\delta$. The DR is operated at a resonance frequency approximately three to nine times higher than that of the QPR, which results in a skin depth roughly two to three times smaller. For illustrative purposes, using the skin depth formula typically applied in the normal skin effect regime (despite its limitations), a $\delta_{\text{QPR}} \approx 0.19 - 0.33$\,\textmu m and $\delta_{\text{DR}} \approx 0.11$\,\textmu m can be estimated when assuming the typical RRR = 100 for OFE copper. This yields a groove depth to skin depth ratio of approximately 105 to 185 for the QPR measurements and 135 to 230 for the DR measurements, despite the deeper grooves of the QPR samples. The higher ratio for the DR measurements could suggest an increased surface resistance than the QPR measurements. Furthermore, the slightly reduced penetration depth in the DR samples may make them slightly more prone to the effects of redeposited particulates. 

These two aforementioned factors may balance each other, leading to potentially similar surface resistance values for both sample types at cryogenic temperature. Lastly, the ratio of groove spacing to spot diameter is 2 for the QPR sample, compared to 0.9 for the DR sample. This suggests differences in groove shapes and angles. The impact of these particular factors on the induced surface current and surface resistance is being theoretically investigated by Madar\'asz \textit{et al.} \cite{Tamas}, where surface roughness models are explored to create a framework describing roughness effects to assist in selecting appropriate treatment parameters.

Now comparing the data acquired in this study, the direct comparison with plain copper reveals the highest surface resistance values relative to the other two studies. The comparison here involves colaminated copper on stainless steel versus bulk copper. Referring again to \Cref{tab:laser1}, it is visible that a different laser was used to prepare the samples. Although the BS samples were prepared with the nominally fastest scanning speed, in the case of 360$^\circ$ spiral, it resulted in grooves that are roughly two to three times deeper than those in the QPR and DR samples, respectively. This difference is likely related to variations in wavelength, pulse length, and repetition rate. Assuming the measured RRR = 80 value for copper of the BS, this leads to a groove depth to effective skin depth ratio of approximately 170 to 335 at 400\,MHz and 1600\,MHz, respectively, which is the highest ratio when compared to the other two studies. In both measurement setups (DR and this study), this treatment, oriented perpendicular to the surface current, resulted in a surface resistance that is equivalent to or greater than that of room temperature copper.

\Cref{fig:DRratio1} compares the results from different studies for the parallel-to-surface-current constellations. As in \Cref{fig:DRratio2}, the entirely treated BS, in this case the 360$^\circ$ longline BS, shows the highest surface resistance compared to the other studies. The data also follows a similar trend to the DR measurement, though notably, the value at cryogenic temperature stands out significantly, being much higher than the other values, suggesting a dominant contribution from its laser-treated surface characteristics. In contrast, the 4 $\times$ 20$^\circ$ longline BS appears to maintain a relatively constant ratio compared to plain copper. The influence of the laser-treated area is less pronounced due to the gradual decrease in induced surface currents horizontally, making it likely that, at room temperature, the required sensitivity falls below the measurement error, where the distinction between treated and untreated surfaces is already minimal. The QPR measurements, on the other hand, exhibit the lowest ratio relative to plain copper.

\begin{figure}[b!]
    \centering
\includegraphics[width=0.9\textwidth]{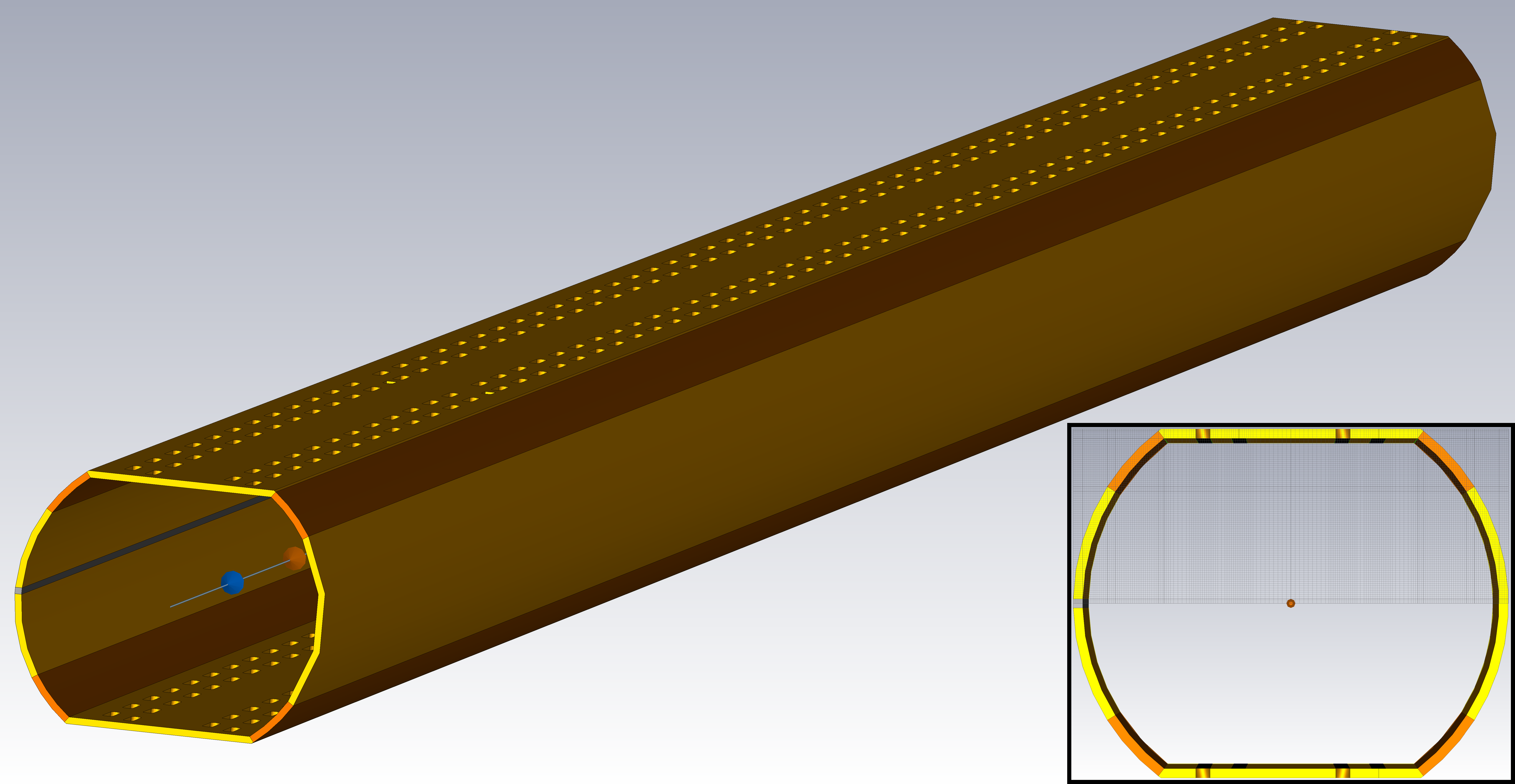}
    \caption{CST wakefield simulation model for the LHC beam screen with 4 $\times$ 20$^\circ$ longline laser treatment configuration. The inset shows the cross-sectional mesh view.}
    \label{fig:CSTBSWake}
\end{figure}

\section{Impact on the LHC - Longitudinal Beam Impedance \label{sec:7}}

The SEY should be low enough to effectively mitigate electron clouds in the quadrupole triplet magnets, with the constraint that the groove depths resulting from laser ablation should not exceed 25\,\textmu m \cite{BezSelective}. This limitation is crucial to avoid a significant increase in the surface impedance of the 75\,\textmu m thick copper layer of the BS. Measurements in this study show that selective laser treatment, specifically the 4 $\times$ 20$^\circ$ pattern, meets these requirements, producing groove depths of 17–22\,\textmu m and a surface resistance increase of only 1.4 times that of untreated copper under operational temperature conditions. However, the locally increased surface resistance in the four treated segments at the corners of the BS may warrant further analysis to assess its potential impact on the real part of the longitudinal beam impedance. Notably, numerical simulations confirm that this configuration effectively mitigates electron cloud formation \cite{BezSelective}.

To evaluate the potential impact of this particular laser treatment on the longitudinal beam impedance of the LHC BS during operation, time-domain simulations were performed using CST's 2024 Wakefield Solver. A one-meter section of the BS was modelled, as shown in \Cref{fig:CSTBSWake}. Solver settings ensured a minimum resolution of 300 mesh cells per wavelength, resulting in approximately 50 million mesh cells per half BS (see inset in the figure). The traversing particle bunch had a root-mean-square (RMS) bunch length of 50\,mm, corresponding to a frequency domain simulation range of up to 2\,GHz. The wake length was set to 100 meters, and all simulations were conducted at 4.2\,K, with particles passing through the centre of the BS.

\begin{figure}[h!]
     \centering
         \includegraphics[width=0.7\textwidth]{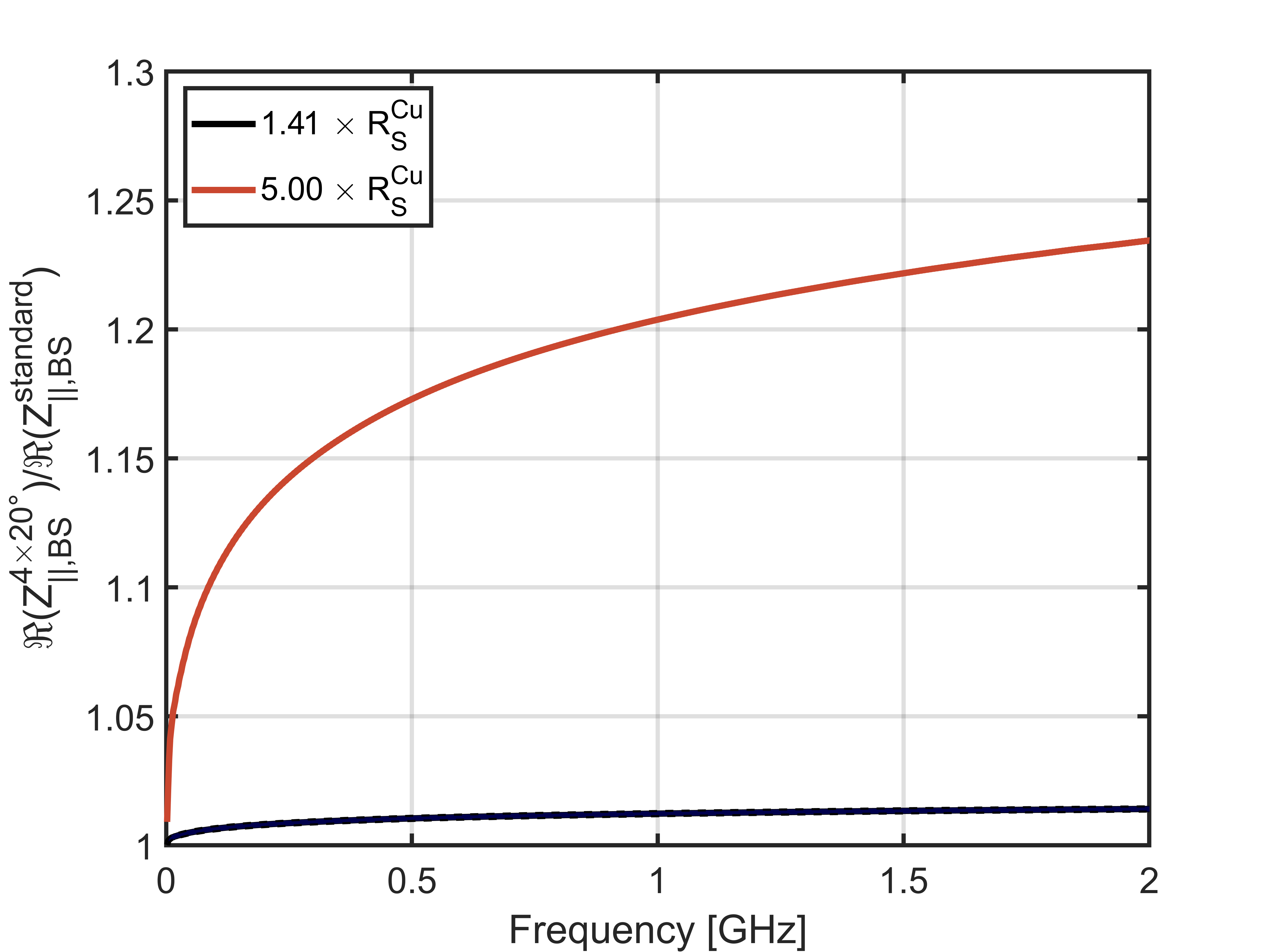}
        \caption{Comparison of the real part of the longitudinal beam impedance: ratio between the 4 $\times$ 20° longline laser-treated LHC beam screen and the standard LHC beam screen.}
        \label{fig:CSTZ}
\end{figure}

\Cref{fig:CSTZ} shows the numerically determined real part of the longitudinal beam impedance as a function of frequency for the measured case, along with a theoretical scenario featuring surface resistance five times greater than that of pure copper under the same treatment area conditions. The results indicate that the real part of the longitudinal impedance, which is entirely broadband, rises by only a minor amount of up to 3\% compared to the standard LHC BS. This slight increase in impedance is primarily attributed to the location of the laser treatment in the four corners of the BS, where the lowest image current density is induced by the circulating beam during operation. Even in the scenario where the treated area exhibits a surface resistance five times higher than pure copper, the increase in impedance would be noticeable by only 25\%.

\begin{figure}[h!]
         \includegraphics[width=0.95\textwidth]{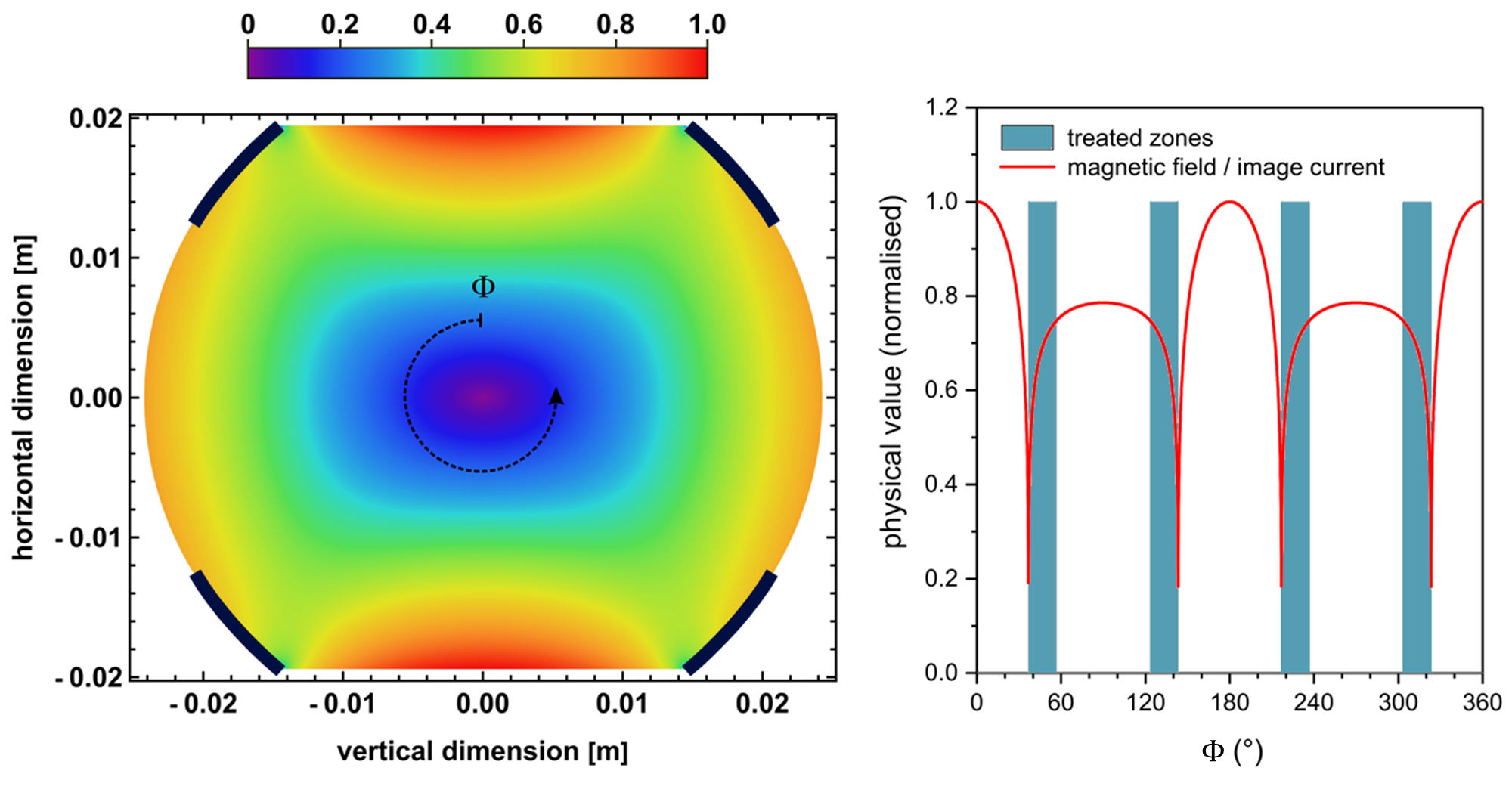}  
        \caption{Left: Density plot illustrating the distribution of normalised magnetic field strength within the LHC beam screens, generated by the circulating particle beam. The dark regions represent the schematically indicated laser-treated areas for the 4\,$\times$\,20° longline pattern. Right: Angular azimuthal dependence of the normalised magnetic field strength at the beam screen surface, along with the corresponding image current. The angle $\Phi = 0^\circ$ is defined at the nearest vertical point to the centre of the beam screen. The laser-treated areas are highlighted in turquoise. Image taken from Bez \textit{et al.} \cite{BezSelective}.}
        \label{fig:BSCurrent}
\end{figure}

This behaviour is further illustrated in \Cref{fig:BSCurrent}, which displays a density map of the normalized magnitude of the magnetic field generated by the particle beam, proportional to the locally induced image current. The angular dependence of these quantities is shown on the right side of the figure, with laser-treated regions marked in grey. The maximum current density is induced in the upper and lower flat sections of the BS, closest to the beam, while the laser-treated areas align with the regions of lowest induced surface current density.

\section{Conclusion \label{sec:8}}
A selective laser treatment process has been developed to suppress electron cloud formation in the quadrupole magnet assemblies of the LHC, aiming to enhance beam quality and reduce heating. For this measurement campaign, four standard LHC beam screens (type 50A) were produced, three of which underwent ultrashort pulse laser treatment to create grooves on their inner surfaces, oriented both parallel and transverse to the beam screen’s longitudinal axis. The feasibility of this method was successfully demonstrated on 40\,cm beam screen segments.

The study revealed that for laser-treated surfaces in accelerator beam screens like those in the LHC, the extent of the treated area, groove depth, and particularly the groove orientation relative to current flow have a significant impact on the equivalent surface resistance of the beam screens.  Additionally, the results have been compared to other studies conducted on flat samples and showed consistency in the determined outcomes. Nevertheless, further investigation into the impact of inclined laser-treated surfaces and stainless steel welding could be performed in the future. 

Surface resistance measurements combined with numerical beam impedance calculations confirmed that the proposed selective laser treatment, utilising a 4 $\times$ 20$^\circ$ pattern, results in no significant increase in the real part of the longitudinal beam impedance compared to untreated copper surfaces. However, the impact on the imaginary part of the beam impedance remains uncertain, as measuring surface reactance is a challenging undertaking. In total, this laser treatment process is a promising solution for implementation during the LHC’s Long Shutdown 3, particularly for specific Q5 standalone magnets near interaction points 1 (ATLAS detector) and 5 (CMS detector).

\section*{Abbreviations}

\noindent{}\begin{tabular}{ll}

ARPE & Algorithm for Resonator Parameter Extraction \\
BS & Beam Screen  \\
CST & Computer Simulation Tool \\
DR & Dielectric Resonator\\
EM & Electromagentic \\
HL & High-Luminosity\\
LESS & Laser-Engineered Surface Structuring \\
LHC  & Large Hadron Collider\\
N & Nitrogen \\
OFE & Oxigen Free Electronic \\
Q-factor & Quality Factor \\
QPR & Quadrupol Resonator \\
RF & Radio Frequency \\
RMS & Root-Mean-Square \\
    \end{tabular}

    \noindent{}\begin{tabular}{ll}

RT & Room Temperature \\
SEY & Secondary Electron Yield \\
SMA & SubMiniature version A \\
VNA &  Vector Network Analyser \\ 
    \end{tabular}
    

  

\addcontentsline{toc}{section}{References}
\bibliography{mybib}

\end{document}